\documentclass{pas}
\usepackage{aas_macros}
\usepackage{pdflscape}

\begin{document}

\lefttitle{Population synthesis of sdBs with COMPAS}
\righttitle{N. Rodriguez-Segovia et al.}

\jnlPage{1}{22}
\jnlDoiYr{2024}
\doival{10.1017/pasa.xxxx.xx}

\articletitt{Research Paper}

\title{Population synthesis of hot-subdwarf B stars with COMPAS: parameter variations and a prescription for hydrogen-rich shells}

\author{\gn{Nicol\'as} \sn{Rodr\'iguez-Segovia}$^{1}$, \gn{Ashley\,J.} \sn{Ruiter}$^{1,2,3}$ and \gn{Ivo\,R.} \sn{Seitenzahl}$^{1}$}

\affil{$^1$School of Science, University of New South Wales, Australian Defence Force Academy, Canberra, ACT 2600, Australia}

\affil{$^2$ARC Centre of Excellence for All-Sky Astrophysics in 3 Dimensions (ASTRO-3D)}

\affil{$^3$OzGrav: The ARC Centre of Excellence for Gravitational Wave Discovery, Hawthorn, VIC 3122, Australia}

\corresp{N. Rodr\'iguez-Segovia, Email: nj.rsegovia@gmail.com}

\citeauth{Rodriguez-Segovia et al., Population synthesis of hot-subdwarf B stars with COMPAS. {\it Publications of the Astronomical Society of Australia} {\bf 00}, 1--22. https://doi.org/10.1017/pasa.xxxx.xx}

\history{(Received xx xx xxxx; revised xx xx xxxx; accepted xx xx xxxx)}

\begin{abstract}

Subdwarf B stars are a well-known class of hot, low-mass stars thought to be formed through interactions in stellar binary systems. While different formation channels for subdwarf B stars have been studied through a binary population synthesis approach, it has also become evident that the characteristics of the found populations depend on the initial set of assumptions that describe the sometimes poorly constrained physical processes, such as common envelope episodes or angular momentum loss during mass transfer events. In this work we present a parameter study of subdwarf B populations, including a novel analytic prescription that approximates the evolution of subdwarf B stars with hydrogen-rich outer shells, an element previously overlooked in rapid binary population synthesis. We find that all studied parameters strongly impact the properties of the population, with the possibility of igniting helium below the expected core-mass value near the tip of the red giant branch strongly affecting the total number of subdwarf B candidates. Critically, our newly proposed prescription for the evolution of subdwarf B stars with hydrogen-shells helps to reconcile theoretical predictions of surface gravity and effective temperature with observational results. Our prescription is useful in the context of rapid binary population synthesis studies and can be applied to other rapid binary population synthesis codes' output.
\end{abstract}

\begin{keywords}
B subdwarf stars, Binary stars
\end{keywords}

\maketitle

\section{Introduction}

Subdwarf B stars (hereafter sdB) are understood as low-mass stars in late stages of evolution, stripped of most of their hydrogen-rich envelope and stably burning helium in their cores while located at the hot end of the horizontal branch of the Hertzsprung-Russell diagram (the so-called extreme horizontal branch). They were first defined by \cite{Sargent1968} and an overview of their properties can be found in the recent review by \cite{Heber2016}. Although formation scenarios from single stellar evolution have been proposed for sdBs \citep[such as the hot flasher case, see][]{DCruz1996}, binary evolution provides the most widely accepted formation channels. This is supported by the large fraction of observed sdBs that are members of a binary system \citep[see e.g.,][and references therein]{Maxted2001,Stark2003,Pelisoli2020}, and by how binary interactions can explain the method through which a star is capable of losing a large fraction of its outer hydrogen-rich layers. Additionally, the existence of single sdBs does not contradict the binary-related formation channels, since single sdBs can also be explained through binary evolution, due to the merger of two helium white dwarfs \citep[He WDs, see i.e.][]{Tutukov1985}. For more discussion about the formation channels, see \citet[][abbreviated as H02 and H03 throughout the paper]{Han2002,Han2003}.

Around one-third of sdBs in the observational sample have short orbital periods \citep{Pelisoli2020} and, within this group, there are many sdB + white dwarf (WD) systems which are thought to lead to energetic transient phenomena such as double detonation type Ia supernovae (see e.g., Nomoto \citeyear{Nomoto1982a}; Livne \citeyear{Livne1990}; Woosley \& Weaver \citeyear{Woosley1994}; Neunteufel et al. \citeyear{Neunteufel2016} for details, or Kupfer et al. \citeyear{Kupfer2022} for a recently proposed double-detonation progenitor) or AM CVn systems \citep[e.g.,][]{Savonije1986,Iben1987,Bauer2021}, when helium-rich material is transferred from the sdB onto the WD. Another important consequence of these small orbits is that such short periods might produce gravitational waves detectable by the upcoming Laser Interferometer Space Antenna \citep[LISA,][]{AmaroSeoane2023}, an example of which is CD–30$^\circ$11223 \citep{Geier2013, Kupfer2018}. 

The importance of sdBs in the context of binary evolution is therefore clear, but it should be noted that their relevance extends further. For example, they have been used in studies of the observed UV upturn of elliptical galaxies \citep[e.g.,][]{Brown2004} and old stellar populations, as well as in the context of asteroseismological studies owing to the different modes in which they have been observed to pulsate \citep[e.g.,][]{Reed2021}. In a similar vein of variable star research, sdBs have also been proposed as potential progenitors of the recently discovered blue large-amplitude pulsator (BLAP) class of pulsating stars \citep{Pietrukowicz2017}, which would correspond to an sdB's helium shell burning stage after helium has been depleted in the core and the star moves towards the subdwarf O star location in the Hertzsprung-Russell diagram \citep[for details, see][]{Xiong2022}.

All of these elements imply that there is a need for a better understanding of the formation and evolution of sdB systems. For this purpose, binary population synthesis (BPS) has been historically used as a statistical tool to study sdB formation channels \citep[e.g.,][]{Han2003, Nelemans2010, Clausen2012, Chen2013}, finding general agreement with observed properties such as the orbital period and mass distribution of the current day population. Despite this, rapid BPS by design requires a large number of parametrizable quantities, making it difficult to set a unique set of input physics that can accurately reproduce the observational results \citep[but see][for a BPS parameter study applied to double WD systems]{Toonen2014}. Instead, it is possible to get an idea of what parameters play a greater role on recovering observed properties of a given population, and also what elements are present in the final populations in multiple different configurations, hinting at a higher likelihood of their occurrence in real populations.

In this paper, we use the rapid BPS code Compact Object Mergers: Population Astrophysics and Statistics \citep[\textsc{compas}, Team COMPAS:][]{Riley2022} and results from Modules for Experiments in Stellar Astrophysics \citep[\textsc{mesa},][]{Paxton2011, Paxton2013, Paxton2015, Paxton2018, Paxton2019} to explore the impact of varying the initial parameters when studying sdBs through binary population synthesis, as well as the effects that these parameters have on the proposed sdB formation channels from binary star evolution. 

Section \ref{sec:method} contains the description of the method and software used in our work, particularly details about our BPS approach and parameter variation in subsection \ref{sec:BPS}, as well as relevant methods related to detailed stellar models in subsection \ref{sec:mesa}. The results are then shown in section \ref{sec:results}, which is organized as follows: the effects of varying parameters in subsection \ref{sec:par_var}, the sdB-formation channels in subsection \ref{sec:channels}, and a comparison to similar studies available in the literature in subsection \ref{sec:comp}. We provide our summary and conclusions in section \ref{sec:conclusion}.

\begin{table}
	\caption{Set of parameters used to construct our different \textsc{compas} population synthesis realizations. A single value (third column) implies that the property was changed from its default value \citep[defined in][]{Riley2022}, but kept fixed on all our runs.}\label{tab:params}
	{\tablefont\begin{tabular}{@{\extracolsep{\fill}}lcl}
		\toprule
		Parameter & Section & Values [Unit]$^{{\rm a}}$ \\
		\hline
        Random Seed & \ref{sec:initparms} & 15\\
        Maximum Evolution Time & \ref{sec:initparms} & 13700 [Myr] \\
        Minimum initial mass & \ref{sec:initparms} & 0.08 [M$_\odot$]\\
        Maximum initial mass & \ref{sec:initparms} & 150 [M$_\odot$]\\
		  $\alpha$ & \ref{sec:CE} &  0.2, 1.0, 1.5\\
        Z & \ref{sec:z} &  0.0012, 0.0142$^{\rm b}$, 0.03\\
        MacLeod Linear Fraction & \ref{sec:angmomlossplusmdot} &  0.0, 0.5, 1.0\\
        Mass Transfer Efficiency & \ref{sec:angmomlossplusmdot} &  0.0, 0.5, 1.0\\
        Mass Transfer Stability & \ref{sec:qcrits} &  $\zeta^{\rm c}$, \texttt{GE20}$^{\rm d}$
		\botrule
	\end{tabular}}
    \begin{tabnote}
{$^{{\rm a}}$ Only when applicable.}\tnp
{$^{{\rm b}}$ Default (solar) value in \textsc{compas}, following \cite{Asplund2009}.}\tnp
{$^{{\rm c}}$ e.g., \cite{Soberman1997} or \cite{Woods2012} for a more recent analysis.}\tnp
{$^{{\rm d}}$ Critical mass ratios as in \citet{Ge2020}, under the adiabatic assumption.}\tnp
\end{tabnote}
\end{table}

\section{Method, software and input physics}\label{sec:method}

\subsection{Binary Population Synthesis}\label{sec:BPS}
To create our binary populations we use \textsc{compas} v02.50.00 \citep[Team \textsc{COMPAS}:][]{Riley2022}, an open source rapid BPS software capable of generating populations of stellar binary systems under a set of parameterized prescriptions that describe the evolution and interaction of their components. Similar to other BPS codes such as \textsc{bse} \citep{Hurley2002}, \textsc{startrack} \citep{Belczynski2008}, \textsc{binary\_c} \citep{Izzard2004}, \textsc{seba} \citep{Toonen2012} or \textsc{cosmic} \citep{Breivik2020}, \textsc{compas} is capable of quickly evolving millions of stellar binary systems in a few CPU hours thanks to its efficient code and simplified prescriptions, allowing statistical studies and tests of the selected physical configuration at the expense of detailed physics like what is done through \mbox{1-D} stellar evolution models produced by software such as \textsc{mesa}, or hydro-dynamical codes \citep[e.g.,][]{Price2018}. We also highlight that, so far, \textsc{compas} has been mostly used to study compact objects: remnants resulting in neutron stars or black holes \citep[e.g.,][]{Broekgaarden2022,Stevenson2022,Wagg2022}. Here we study sdBs (which have masses below $1~M_\odot$, typically around $0.5~M_\odot$), effectively using \textsc{compas} to study low-mass non-compact remnants for the first time. Our results will thus set the stage for future \textsc{compas} research involving stars and their products at the low- and intermediate-mass end of the initial mass function. 

To analyse the different sdB formation pathways, similar to what has been done in previous studies \citep[e.g., H03;][]{Clausen2012}, we perform a parameter study and we use a total of 162 different combinations of parameters (see Appendix \ref{app:runs} for details of each). Most configurable parameters in \textsc{compas} have not been modified, except for those directly specified in Table \ref{tab:params}. All of the parameters included in this table are discussed in the following sections.

\begin{table}[]
\caption{Summary of the \citet{Hurley2000} stellar types. For simplicity, we use the abbreviations through the text and figures.}\label{tab:sttypes}
{\tablefont\begin{tabular}{@{\extracolsep{\fill}}lll}
\toprule
Type & Abbreviation & Meaning \\
\hline
0  & MS$\leq 0.7$    &  Main sequence star, mass equal or lower than $0.7~M_\odot$ \\
1  & MS$> 0.7$    &  Main sequence star, mass higher than $0.7~M_\odot$ \\
2  & HG    &  Star crossing the Hertzsprung gap\\
3  & FGB   &  First giant branch star\\
4  & CHeB  &  Core helium burning star\\
5  & EAGB  &  Early asymptotic giant branch star\\
6  & TPAGB &  Star in the thermally pulsing stage of the\\
 & & asymptotic giant branch\\
7  & HeMS  &  Helium main sequence star\\
8  & HeHG  &  Star crossing the helium Hertzsprung gap\\
9  & HeGB  &  Helium giant branch star\\
10 & HeWD  &  Helium white dwarf\\
11 & COWD  &  Carbon-oxygen white dwarf\\
12 & ONeWD &  Oxygen-neon white dwarf\\
13 & NS    &  Neutron star\\
14 & BH    &  Black hole\\
15 & MR    & Massless remnant
\botrule
\end{tabular}}
\end{table}

Our sample of sdB candidates is built by collecting stars flagged as type 7 at any given time during their evolution, that is, they have gone through the naked helium main sequence (HeMS) as per the \textsc{compas} notation for stellar types \citep[see Table \ref{tab:sttypes} for the definition of each stellar type, inherited from][]{Hurley2000}. On top of that, we also require our sample to match or be approximately equal to the observational results (as seen in Fig. \ref{fig:kiel_obs}), effectively imposing a restriction on the size and temperature of each candidate. This criterion is enforced by selecting HeMS stars that cross the area in the Kiel diagram delimited by the following equations:

\begin{align}
    \log{g} = 6\log{T} - 20.4,\\
    \log{g} = 6\log{T} - 22.4,\\
    \log{g} = -3.4\log{T} + 21.8,\\
    \log{g} = -3.4\log{T} + 19.3,
\end{align}

\noindent where $T$ is the effective (in Kelvins) and $g$ surface gravity (in cm s$^-2$) of a given star. This has been defined by visual inspection of the \citet{Culpan2022} observational sdB sample, and further checked against the \citet{Lei2023a} sdB catalog as seen in Fig. \ref{fig:kiel_obs}. Note that this method allows for the existence of stars that are not necessarily born, or stay during their entire lifetimes, as our observational definition of sdB candidates.

We also select some candidates from stars flagged as type 10 (helium white dwarfs, HeWD), only when we want to analyse the impact of allowing helium ignition in HeWDs with masses within 5\% or 3\% of the expected core mass the progenitor would have attained at the tip of the red giant branch (RGB, used as a synonym of FGB defined in Table \ref{tab:sttypes}), owing to the results of \citet{DCruz1996}, H02 and \citet{Clausen2012}. These stars are then evolved as HeMS stars during post processing, and follow the same selection criteria based on observational constraints as HeMS stars do.

A final important consideration is that within our candidates we do not include systems that merge right after a mass transfer event, nor do we track the evolution of the companions. This last element is a consequence of our post-processing approach to the evolution of the sdB candidates, as will be explained in section \ref{sec:h_prescription}.

\subsubsection{Initial Parameters}\label{sec:initparms}

Both initial masses and orbital configuration (orbital periods and eccentricities) for all our systems are sampled following the correlated distributions presented in \cite{Moe2017}. This process is implemented in the \texttt{sampleMoeDiStefano.py} script distributed as part of the pre-processing tools in \textsc{compas}. The only parameter that we have customized in this script is the mass range, which we have set to 0.08 - 150$\, M_\odot$ for the initial mass of the primary star to allow all possible progenitors: non-zero accretion efficiency allows low-mass stars to evolve faster (eventually becoming sdB candidates) as a consequence of an accretion episode, while setting a rather high upper mass limit allows for a higher chance of sampling massive primaries. About 1,000,000 \textit{stellar systems} are produced, though this also includes single stars that are not evolved by \textsc{compas} when using its binary population synthesis mode. After removing these, our binary population consists of 
$\sim 284,000$ systems, which will be evolved using the 162 different configurations previously mentioned.

\subsubsection{Common Envelope}\label{sec:CE}

The \textsc{compas} code adopts the energy formalism for common envelope \citep[CE,][]{Webbink1984, deKool1990}, which requires setting the $\alpha$ and $\lambda$ parameters as indicated by the equations

\begin{align}
    E_{\rm{bind}} = \alpha \, \Delta E_{\rm{orb}},\label{eq:alpha}\\
    E_{\rm{bind}} = - \frac{GMM_{\rm{env}}}{\lambda R},
\end{align}

\noindent where $E_{\rm{bind}}$ is the gravitational binding energy of the envelope, $\alpha$ is an efficiency parameter that specifies the fraction of orbital energy used to remove the CE, $\Delta E_{\rm{orb}}$ corresponds to the change in orbital energy due to the CE phase, and $\lambda$ is a structure parameter that represents the relationship of the binding energy of the stellar envelope and the location of its inner boundary, as well as the sources of energy considered for its removal (for a recent review on numerical techniques relevant for CE evolution, see \citeauthor{Roepke2023} \citeyear{Roepke2023}). 

For $\lambda$ we use the default configuration in \textsc{compas}, i.e. the \cite{Xu2010, Xu2010a} prescription implemented through fitting formulae to results of detailed stellar models, considering the full contribution of internal energy \citep{Riley2022}. In the case of $\alpha$ we explore the set of values presented in Table \ref{tab:params}, as the current constraints on envelope removal efficiency are poor \citep{DeMarco2011}, and therefore it is useful to explore different possible values. \cite{Zorotovic2010} point towards $\alpha_{\rm{CE}} \sim 0.2$, while we also consider full efficiency ($\alpha = 1$) and the possible contribution of additional energy sources with $\alpha = 1.5$ \citep[see e.g.,][]{Ivanova2020}.

\subsubsection{Metallicity}\label{sec:z}

The importance of metallicity in the formation of sdBs can be understood from a few different angles. First, changes on metallicity modify the initial mass -- core mass relation close to the tip of the RGB \citep[see e.g.,][]{Cassisi2016}, affecting the observed sdB mass distribution by changing the core mass value near the tip of the RGB. This core mass is the one that directly sets the mass of a given sdB, as it represents most of what remains of the progenitor star after removing its outer layers. It must be noted that other parameters such as the overshooting prescription might play an important role in setting the initial mass -- core mass relation as well \citep[e.g.,][and references therein]{ArancibiaRojas2024}. Next, the mass limit for the ignition of helium in a flash (MHeF) is also affected by changes on metallicity. \cite{Sweigart1978} show that the critical mass threshold for a helium flash to occur is lowered with decreasing metallicity. This affects the initial mass and proprieties of the newly born sdB, particularly in the context of rapid BPS codes that have inherited the \citet{Hurley2000} methods where the evolution of a HeMS star (used as an sdB proxy) depends solely on its initial mass. Note that while the critical mass is reduced, the core mass near the tip of the RGB for a given zero age main sequence (ZAMS) mass value is increased. Finally, the size of the progenitor (the donor) impacts the likelihood of Roche Lobe Overflow (RLOF) and therefore mass transfer, as well as its stability \citep[e.g.,][]{Chen2013, Vos2020}. The role played by metallicity is relevant, though the growth of the stellar radius during the RGB phase is still a topic of discussion  \citep[e.g.,][]{Renzini2023}.

To explore its effects on the final sdB yields, our BPS realizations include 
three different metallicities, namely sub-solar, solar, and super-solar. The specific values were chosen considering both \textsc{compas} limits and the study of long period sdBs in \cite{Vos2020}, which considers the different structural components of the Galaxy. Their metallicities are shown as iron abundance relative to hydrogen (i.e., [Fe/H]) and are closer to a continuous distribution, while we use a discrete set of 3 values (see Table \ref{tab:params}) and relate [Fe/H] to nominal Z input values for \textsc{compas}. This was done by using a simplified approach \citep[e.g., Equation 9 in][assuming that $X\approx X_\odot$]{Bertelli1994}:

\begin{equation}
    \left[\rm{Fe/H}\right] \approx \log{Z/Z_\odot},
\end{equation}

\noindent which approximately transforms the range $-1.08 \leq \left[\rm{Fe/H}\right] \leq 0.4 $ covered in table 1 of \cite{Vos2020}  to $0.0012 \leq Z \leq 0.036$, by assuming $Z_\odot = 0.0142$ \citep{Asplund2009} as implemented in \textsc{compas} \citep[Team COMPAS:][]{Riley2022}. The final values that we use consider these results (limited by the maximum/minimum allowed metallicity value in \textsc{compas}), as well as solar metallicity. We have ignored the metallicity value related to the halo component, as it does not seem to play a critical role for the Galactic sdB population (consider its mass fraction in table 1 of \citeauthor{Vos2020} \citeyear{Vos2020}).

\subsubsection{Orbital Angular Momentum Loss And Mass Transfer Efficiency}\label{sec:angmomlossplusmdot}

We assume that the specific angular momentum carried away by any amount of mass being lost from the system is a fraction of the total specific angular momentum, represented by

\begin{align}
    h_{\rm{lost}} = \gamma \frac{J}{M_{\rm a} + M_{\rm d}},\label{eq:hlost}\\
    J = M_{\rm a}M_{\rm d}\sqrt{\frac{Ga\left(1-e^2\right)}{\left(M_{\rm a} + M_{\rm d}\right)}},\label{eq:jorbtot}
\end{align}

\noindent with $h_{\rm{lost}}$ the specific angular momentum being lost, $J$ the total orbital angular momentum of the system, $G$ the gravitational constant, $a$ the semi-major axis, $e$ the eccentricity and $M_{\rm x}$ is the mass of the donor (d) or accretor (a). Note that $e$ can be taken as 0 here, since \textsc{compas} circularizes orbits right before mass transfer events (a common practice in rapid BPS codes). $\gamma$ represents a coefficient that specifies the fraction of specific angular momentum being lost, and it depends on the position at which mass is being lost from the system. This can be expressed as

\begin{align}
    h_{\rm{lost}} = a_{\rm{lost}}^2\omega,\label{eq:hobj}\\
    \omega \equiv \frac{2\pi}{P} = \sqrt{\frac{G\left(M_{\rm a} + M_{\rm d}\right)}{a^3}},
\end{align}

\noindent where $\omega$ is the orbital frequency and $P$ its orbital period. All other elements are the same as in equation \ref{eq:jorbtot}. Then, by combining equations \ref{eq:hlost}, \ref{eq:jorbtot} and \ref{eq:hobj} we can find an expression for $\gamma$. Explicitly,

\begin{equation}
    \gamma = \left(\frac{a_{\rm lost}}{a}\right)^2 \frac{\left(M_{\rm a} + M_{\rm d}\right)^2}{M_{\rm a}M_{\rm d}\sqrt{1-e^2}},\label{eq:gamma}
\end{equation}

\noindent which shows the dependence on where mass is being lost from (by setting the appropriate $a_{\rm{lost}}$ value).

This framework has been implemented within the \texttt{MACLEOD\_LINEAR} prescription in \textsc{compas} \citep{Willcox2023}, where instead of setting $a_{\rm lost}$ we can set the linear variable \texttt{--mass-transfer-jloss-macleod-linear-fraction}\footnote{Older versions of \textsc{compas} follow this notation, while newer versions of the code allow for different values depending on whether the accretor is degenerate or not. We have used a single value for both accretor types.} (MLF; MacLeod Linear Fraction) with values between $0$ and $1$, corresponding to $a_{\rm{obj}} \in \left[a_{\rm{acc}}, L_2\right]$ (the position of the accretor and the second Lagrangian point, respectively). We choose 3 different configurations: mass is lost from the accretor (usually called isotropic re-emission), from the middle point between the accretor and L2, or from L2.

As for the case of mass transfer, we use a fixed mass accretion efficiency, with possible fractional values $\beta$ corresponding to 0, 0.5, 1. Explicitly:
\begin{equation}
    \dot{M}_{\rm d} = -\beta\dot{M}_{\rm a },\label{eq:mteff}
\end{equation}

\noindent with $\dot{M}$ the mass change rate (sub-indices are the same as in previous equations). The chosen values cover both extreme cases, full and no accretion; as well as an intermediate scenario.

The final effect on the orbit's semi-major axis can be found by taking the time derivative of equation \ref{eq:jorbtot}, coupled with equations \ref{eq:gamma} and \ref{eq:mteff} to simplify the resulting expression. This yields

\begin{equation}
    \frac{\dot{a}}{a} = -2\frac{\dot{\rm{M}}_{\rm d}}{{\rm{M}}_{\rm d}}\left[1-\beta\frac{M_{\rm d}}{M_{\rm a}}-\left(1-\beta\right)\left(\gamma+0.5\right)\frac{M_{\rm d}}{M_{\rm d}+M_{\rm a}}\right],\label{eq:axis_change}
\end{equation}

\noindent where we can see that the change in the orbit's semi-major axis during a mass transfer event depends on the masses of the components, the donor's mass change rate, the accretion efficiency ($\beta$) and the location at which mass is being lost from the system ($\gamma$)\footnote{For more details, the reader is referred to Onno Pol's lecture notes on binary stars, chapter 7.2.2: \href{https://www.astro.ru.nl/~onnop/education/binaries_utrecht_notes/}{https://www.astro.ru.nl/$\sim$onnop/education/binaries\_utrecht\_notes/}}.

We also remark that even though we have focused in the effects of mass transfer accretion efficiency on orbital evolution, a non-zero value would result in mass gain for the accretor, which in turn has its evolution modified. See \citet{Hurley2000} and \citet{Riley2022} for details.

\begin{figure}[h]
\centering
\includegraphics[width=1\linewidth]{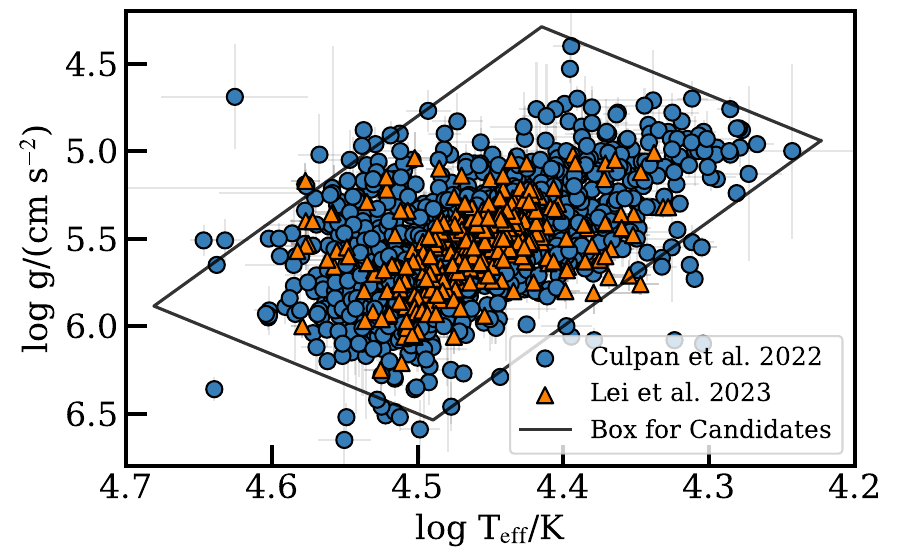}
\caption{Kiel diagram for a sample of known sdBs, where blue circles correspond to \citet{Culpan2022} and orange triangles to \citet{Lei2023a}. Their reported uncertainties are shown as light grey lines. The box delimited by black solid lines corresponds to the selection criteria chosen as our definition of sdB candidates from the \textsc{compas} sample and is explicitly defined in section \ref{sec:BPS}.}
\label{fig:kiel_obs}
\end{figure}

\subsubsection{Mass Transfer Stability}\label{sec:qcrits}

Once a star overflows its Roche lobe, whether the ensuing mass transfer is dynamically stable or not (leading to a CE event) is considered to depend on the response of both its radius and Roche lobe radius to mass loss \citep{Paczynski1972,Hjellming1987,Soberman1997}. In \textsc{compas}, the default approach to evaluate this stability is to use the $\zeta$ prescription \citep[e.g.,][]{Soberman1997} as presented in section 4.2.2 of \cite{Riley2022}. Additionally, we include the \cite{Ge2020} critical mass ratios (under the adiabatic assumption) as an alternative in order to test the effect of using a different, more recent prescription when evaluating the stability of a mass transfer event.

Evidently, we expect different mass transfer stability prescriptions to return different ratios of candidates being born from the stable mass transfer channel to the CE-related channels. This in turn should impact both the period distribution and number of mergers, whether they lead to an sdB candidate or not. Consequently, the total yield of sdBs would be modified as well.

\subsection{Additional elements from MESA detailed models}\label{sec:mesa}

\subsubsection{Hydrogen-rich Layer}\label{sec:h_prescription}

The presence of an outer thin hydrogen-rich layer \citep[see e.g.,][and references therein]{Brassard2001, Krzesinski2014, Hall2016} in sdBs seems to play an important role as a regulator of both surface temperature and size, as depicted by figure 2 in H02, or figures 1 and 2 in \cite{Bauer2021}. When there is little to no envelope, detailed stellar structure models show that a given sdB could be much more compact and hotter than a different sdB of similar mass possessing a $\sim10^{-3}\,\rm{M}_\odot$ hydrogen-rich layer, a difference that would evidently affect observables such as surface gravity and effective temperature. This can be seen in Fig. \ref{fig:bauer_models} where $\Delta T \sim 10,000~K$ and $\Delta\log{g}\sim0.5$ in the most extreme cases.

The current stellar evolution prescription in \textsc{compas}, and most BPS codes that have adopted the \cite{Hurley2000} scheme for stellar evolution, is limited to phases of \textit{naked} helium stars only, which means that no hydrogen-rich envelopes have been considered for them. To get results closer to the observed sample of sdBs shown in Fig. \ref{fig:kiel_obs}, we adopt models built using \textsc{mesa} as presented in \cite{Bauer2021}. These models include sdBs of masses below $0.58 M_\odot$ with ($10^{-3}$ or $3\times10^{-3} M_\odot$) and without hydrogen-rich envelopes. We refer the reader to that work for technical details, as in this paper we focus solely on the prescription developed to incorporate their results into our sdB population synthesis scheme.

To create a new rapid BPS-friendly prescription, we start by taking the same approach as \cite{Hurley2000} and look for maximum sdB age (defined as the time spent core-helium burning) as a function of helium zero age main sequence (ZAHeMS) mass, which corresponds to the mass value as soon as the sdB progenitor has been stripped of most of its outer hydrogen-rich mass. Then, both radius and luminosity are taken as a function of relative age (time elapsed since ZAHeMS normalized by maximum age, that is, values between zero and one), hydrogen-rich shell mass and ZAHeMS mass. We note that, unlike models without hydrogen-rich shells, there is an additional dependence on the ZAMS mass of the progenitor: the models behave differently for stars with ZAMS below MHeF when compared to those above this critical mass limit, which ignite helium smoothly. An example of this behavior is shown in Fig. \ref{fig:bauer_models}. This creates an additional challenge: since the value for MHeF changes between different stellar models and is affected by several parameters \citep[e.g., metallicity, overshooting, and more;][]{Cassisi2014, Ghasemi2017}, its exact value is not the same for \cite{Hurley2000} and \cite{Bauer2021}, showing that an ideal prescription for the evolution of HeMS stars with hydrogen-rich shells should also depend on the MHeF value predicted by the specific set of detailed models being analyzed. Nonetheless, we proceed assuming that we can use the results presented in \cite{Bauer2021} without expanding the grid of models to account for the discrepancies, and apply them to the stars evolved through \cite{Hurley2000} prescriptions as a first approach to the hydrogen envelope problem. However, we do consider the MHeF values computed following \cite{Hurley2000} to decide whether we apply the prescription for the evolution of HeMS with hydrogen-rich shells predicted by the models that experience smooth helium ignition, or the one predicted by models that experience ignition in a flash instead.

After visual inspection of the different parameters involved in the models and attempting several different possible mathematical descriptions, we propose the following equations as a simplified prescription for the evolution of HeMS stars with hydrogen-rich shells:

\begin{figure}[h]
\centering
\includegraphics[width=1\linewidth]{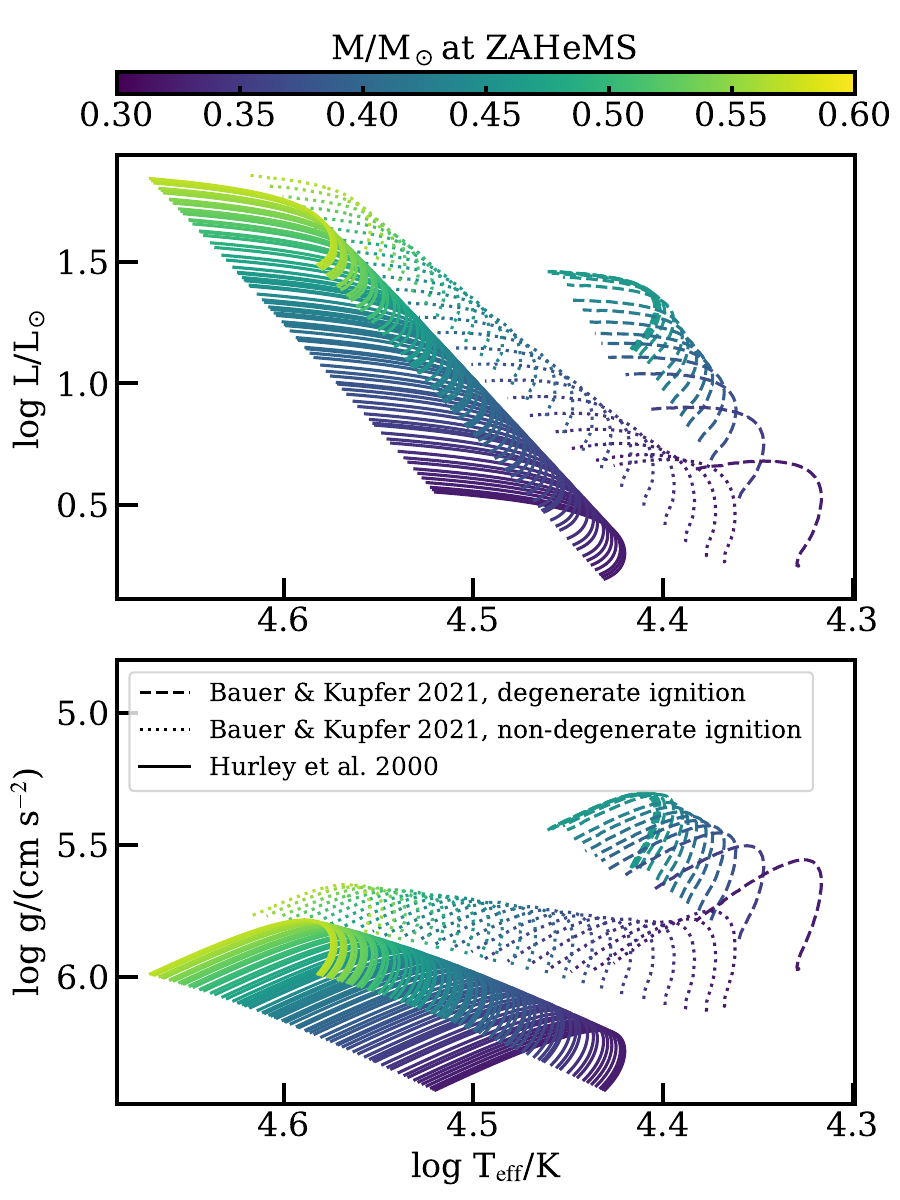}
\caption{Hertzsprung-Russell (top) and Kiel (bottom) diagram for HeMS evolutionary tracks of different mass values. Note that the \citet{Hurley2000} models do not consider any hydrogen rich outer layers, while the \citet{Bauer2021} models shown here consider a $10^{-3}$ M$_\odot$ hydrogen rich outer layer. The difference between stars that ignited helium in a flash (degenerate) and those that did it smoothly (non-degenerate) is noticeable.}
\label{fig:bauer_models}
\end{figure}

\begin{align}
    \tau_{\rm He} = \left(\frac{A_1 - M_{\rm H}}{M + M_{\rm H} - A_2} - A_3 - M\right)\frac{1}{A_4 M} + A_5,\label{eq:lifetime} \\
    t_{\rm r} = \frac{t}{\tau_{\rm He}},\label{eq:rel_time} \\
    R = B_1 + B_2t_{\rm r} - {\rm exp}\left(\left(1+M_{\rm H}\right)\left(B_3+B_4t_{\rm r}\right)\right),\label{eq:radius} \\
    L = C_1(t_{\rm r} + C_2) + C_3(t_{\rm r} + C_4)^2 - 1000\frac{(t_{\rm r} + C_5)^{10}}{1 + 25M_{\rm H}} + \frac{1.5}{1 + C_6t_{\rm r}},\label{eq:luminosity}
\end{align}

\noindent with $\tau_{\rm He}$ the total time in Myr spent in the HeMS phase\footnote{In these \textsc{mesa} models, the HeMS phase lasts until the core helium-4 mass fraction drops below 1\%}, $t$ the time in Myr ellapsed since the ZAHeMS, $M$ mass (no subscript indicates total mass, while an H subscript represents the hydrogen-rich shell mass), $R$ radius and $L$ luminosity. These last three elements are defined using solar units. The coefficients ($A_i, B_i, C_i$ with $i$ a positive integer) were computed by making use of the \textsc{scipy} \citep{Virtanen2020} method \texttt{curve\_fit}. Each of these coefficients is defined in appendix \ref{app:coefs} as a function of mass at ZAHeMS, mass at ZAMS, and hydrogen-rich shell mass. Overall, this prescription closely follows the stellar evolution predicted by the detailed models, as can be seen in Fig. \ref{fig:bauer_diff}. 

As a final remark, we stress that this prescription is based on models constructed at solar metallicity, has not been tested for extrapolation \citep[potentially generating results that are not reliable for parameters outside the range covered in][]{Bauer2021}, and has not been directly incorporated into \textsc{compas} either. Consequently, all of the results shown in the following sections have been acquired through post-processing of the initial \textsc{compas} products.

\begin{figure}[h]
\centering
\includegraphics[width=1\linewidth]{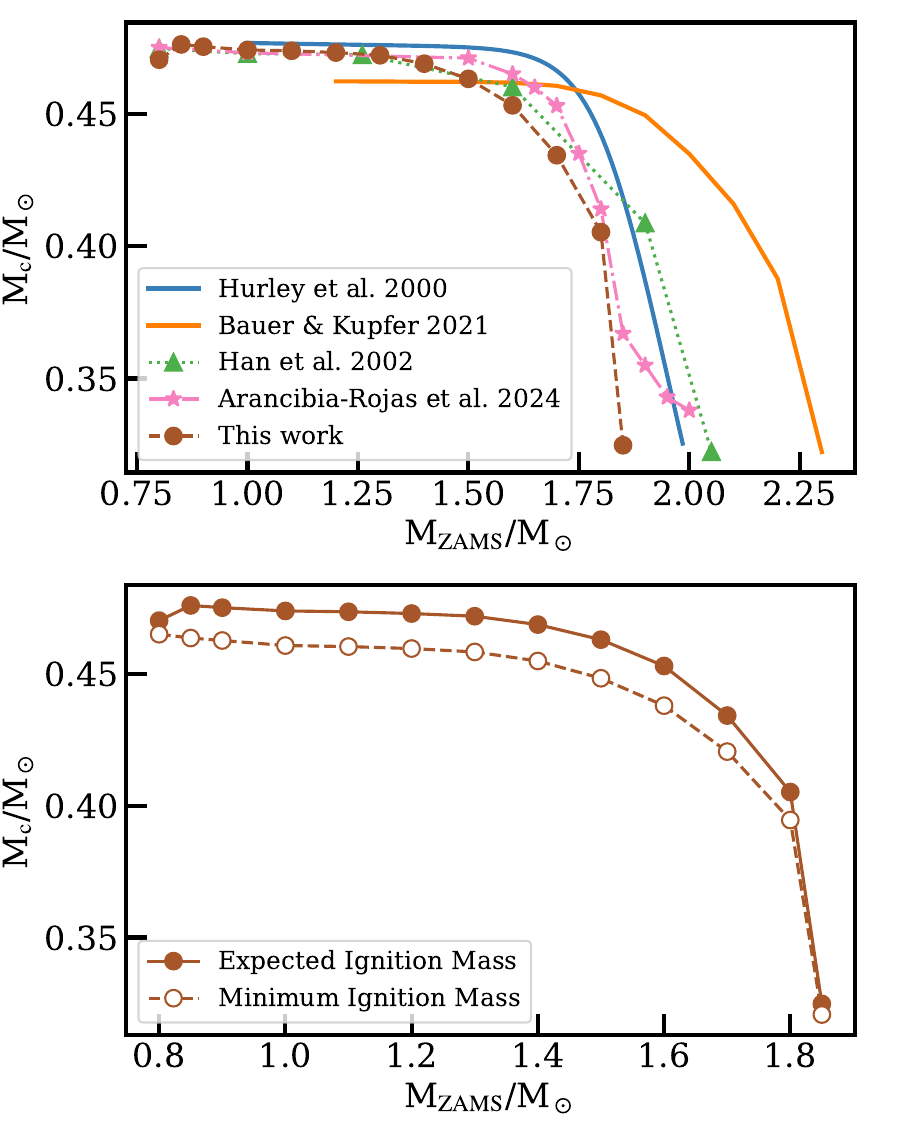}
\caption{The top panel shows how different methods predict the helium core mass (M$_{\rm c}$) at helium ignition as a function of the mass at ZAMS, for models that experience a helium flash and have solar metallicity (defined as $Z = 0.02$). Note that only the \citet{Bauer2021} models do not incorporate any overshooting prescription. On the bottom panel, a zoom-in of our \textsc{mesa} models (filled circles) is shown, alongside the minimum mass at which they were able to experience helium ignition (empty circles, explained in section \ref{sec:corehe}).}
\label{fig:core_range_and_mass}
\end{figure}

\subsubsection{Minimum Core Mass for Helium Ignition}\label{sec:corehe}

An additional element to explore is the possibility of helium ignition while the helium core mass is still below the expected value at the tip of the RGB for stars with masses below MHeF, as explored in H02 \citep[motivated by ][]{DCruz1996}. Inspired by the role that this parameter could potentially play in recovering the canonical mass distribution of sdBs and their birth rate, we use \textsc{mesa} to compare with the results presented in H02. We note, however, that the usage of different software and input physics generate noticeable differences between models \citep[e.g.,][]{Ostrowski2021}. A clear example of this issue is shown in the top panel of Fig. \ref{fig:core_range_and_mass}.

Our simplified approach consists of using \textsc{mesa} v21.12.1, specifically two modified versions of the included \texttt{1M\_pre\_ms\_to\_wd} test suite: one in which evolution proceeds until right after helium is ignited in the core, from which we record the expected core mass at helium ignition; and another in which we stop evolution once the core has reached a given percent of the expected core mass at ignition, typically 95\%. Afterwards, we artificially remove the envelope by relaxing the mass until only the helium core remains, at a mass loss rate given by $\dot{M} = 10^{-3} \frac{M_\odot}{\rm{yr}}$. Then, we let the star continue evolving until the core is cold enough (below 4000$K$, which implies evolution towards a helium white dwarf) or it ignites helium burning. By using a bisection method we can find the minimum percentage at which helium still ignites. Masses in the range of 0.8-1.85 M$_\odot$ were tested (Fig. \ref{fig:core_range_and_mass}, bottom panel), yielding an overall 3\% difference from the expected core mass at the helium flash. This represents a 2\% difference from typical 5\% value found in H02.

It is important to highlight that the percentage differences between H02 and our results can be increased or decreased by tuning the parameters of the model, so these discrepancies should not be considered to be conclusive on their own. Particularly, we neglected the hydrogen-rich shells, as we wanted to check what would the threshold be for naked helium stars, such as the ones that the \citet{Hurley2000} prescription is based on. Additionally, in our implementation the parameters required for the mixing length and overshooting prescriptions (namely \texttt{mixing\_length\_alpha} and \texttt{overshoot\_f0}) were not selected in order to minimize differences with H02, but computed through the solar simplex calibration included in \textsc{mesa}. We also note that our results could imply that the percentage differences are caused by a dependence of the minimum mass for helium ignition on the mass of the hydrogen-rich shell.

In a more detailed study, \cite{ArancibiaRojas2024} have recently explored the helium core mass ignition range for a higher number of initial masses and two different metallicity values (solar and subsolar), finding overall agreement with the 5\% range presented in H02. An important difference, however, is that the 5\% limit is relaxed for stars with progenitor masses above MHeF (those that ignite helium smoothly).

\section{Results}\label{sec:results}

\subsection{Impact of Parameter Variations}\label{sec:par_var}
\begin{figure}[h]
\centering
\includegraphics[width=1\linewidth]{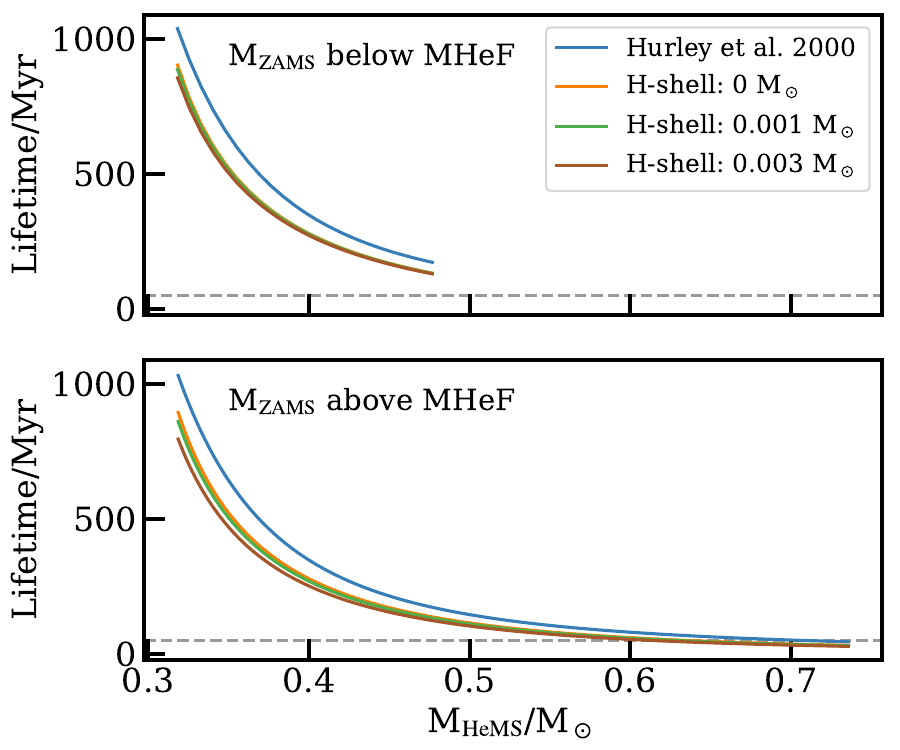}
\caption{Maximum time spent as a HeMS star as a function of mass. The top panel shows stars that experience a flash at helium ignition, while the bottom panel depicts the smooth ignition scenario. The horizontal dashed line shows a lifetime equal 50 Myr, highlighting that candidates with masses above $\sim\!0.58\,{\rm M}_\odot$ spend a comparatively short time as HeMS stars.}
\label{fig:lifetimes}
\end{figure}

The analysis in the following subsections is performed by taking the results from all 162 different configurations, and then dividing them in sub-groups that share characteristics in common. For example, when analyzing metallicity, 3 groups are found (subsolar, solar and supersolar) and compared against each other. The overall number of sdB candidates from the 162 configurations corresponds to 486,425 systems, where only those binaries in which at least one member is classified as a HeMS star that falls within the selected space in the Kiel diagram (Fig. \ref{fig:kiel_obs}) has been considered. We also impose a cut on mass, limiting ourselves to sdB candidates with $M \leq 0.58~M_\odot$ due to the constraints imposed by the range of masses covered by the \cite{Bauer2021} models. Although this affects the total number of systems that we recover, we note that candidates above this mass limit would quickly evolve and not spend much time as sdBs, as shown in Fig. \ref{fig:lifetimes} and also implied in both \citet{Yungelson2008} and \citet{ArancibiaRojas2024}. Similarly, their more massive progenitors (only smooth helium ignition can create such massive HeMS stars)  
correspond to rather low formation probabilities when compared to sdB candidates with masses below $\sim 0.5~M_\odot$ that are born from progenitors with masses under MHeF, due to the the initial mass function.

\subsubsection{Common Envelope: The $\alpha$ Efficiency Parameter}

\begin{figure*}[h]
\centering
\includegraphics[width=1\linewidth]{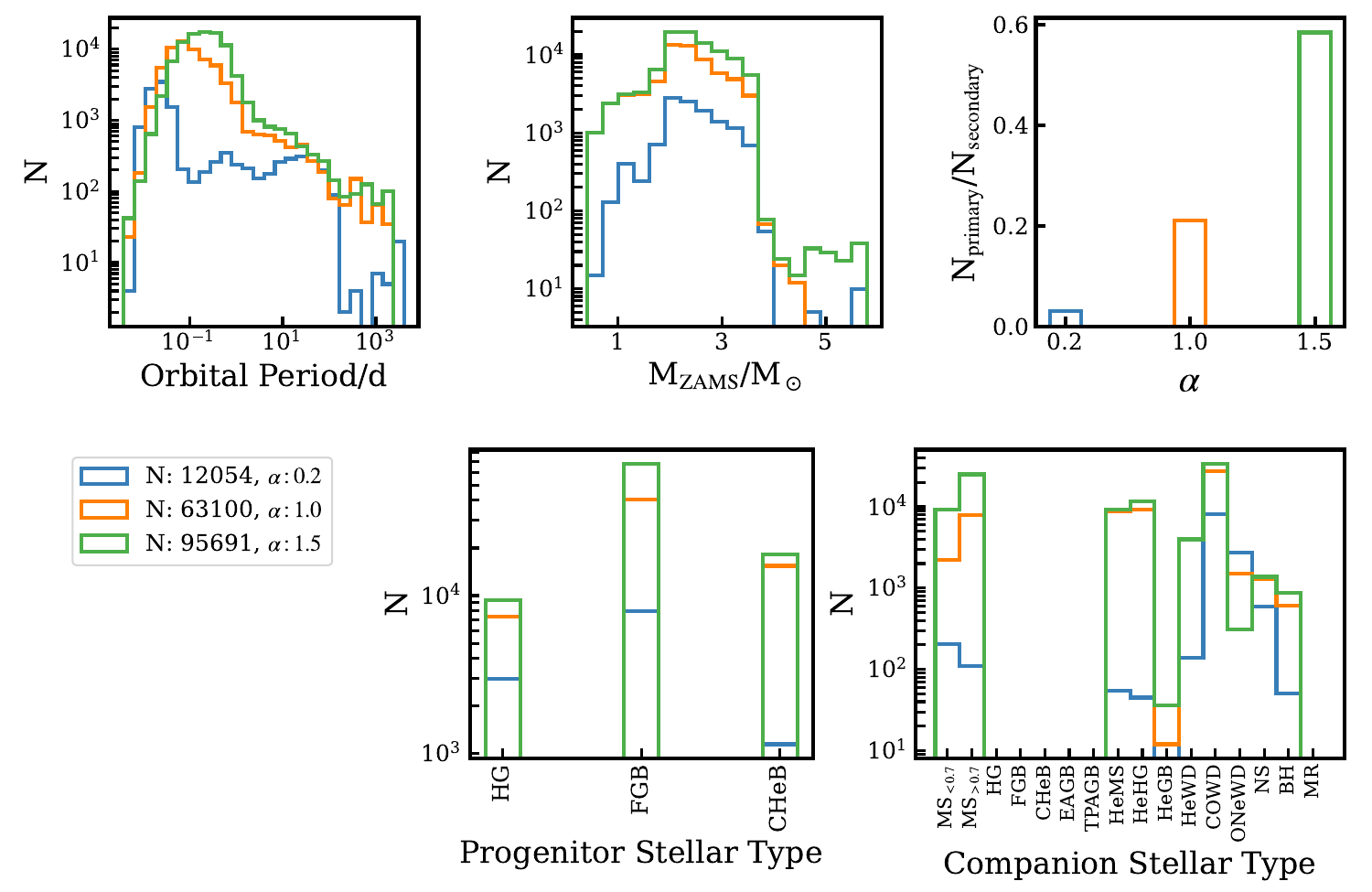}
\caption{Top row, from left to right: Number of candidates per logarithmic-period bin, ZAMS mass distribution, and ratio of candidates being born from the initially primary star to the initially secondary star. Bottom row, from left to right: Stellar type of the progenitor and stellar type of the companion at the primary's ZAHeMS stage (stellar types are defined in table \ref{tab:sttypes}). Note that we present sub-samples grouped by $\alpha$ value, in order to highlight the effects of modifying this parameter. The blue, orange and green histograms correspond to $\alpha$ values equal to 0.2, 1.0, and 1.5, respectively. Only systems that have experienced at least one CE episode are shown.}
\label{fig:alphas}
\end{figure*}

The analysis of this parameter only includes systems that have experienced at least one CE event, as otherwise changes in $\alpha$ would have no impact on the binary stellar population. By considering its definition from Equation \ref{eq:alpha}, we expect lower $\alpha$ values to result in less systems surviving CE episodes due to more frequent mergers within the population, as the higher amount of orbital energy required to eject the envelope of the potential HeMS progenitor star would result in smaller separations after the CE. This can be seen in Fig. \ref{fig:alphas}, where we show all systems that have undergone at least one CE episode before becoming a HeMS star. As expected, lower alpha values result in fewer systems becoming HeMS, with the total number of the $\alpha = 0.2$ configuration being about 13$\%$ of the number of systems found for $\alpha = 1.5$. Despite the differences in total numbers, most systems that evolve up to the HeMS stage through any channel involving at least one CE episode are typically found in short period systems, below $\sim 10$ days and peaking between 0.1 -- 1 days in the logarithmic period distribution. An exception to this property is seen for $\alpha = 0.2$, due to the low efficiency and consequent higher amount of orbital energy spent in removing the envelope of the sdB progenitor, resulting in more compact orbits and a peak below 0.1 days in the logarithmic period distribution. This effect also explains the slight differences in the location of the logarithmic period distribution peak for the remaining $\alpha$ values. We also note that long period systems are found in these CE channels, though most of them can be understood as systems that experienced a CE episode at an early evolutionary stage, and therefore was not the direct precursor of the HeMS star. A latter stable mass transfer would then form the sdB candidate and also increase the orbital separation.

Interestingly, the increasing number of candidates for higher $\alpha$ values seems to be linked to progenitors with ZAMS masses above MHeF that eventually evolve into HeMS stars. As for the companions of our candidates, there are no clear trends linked to the changing $\alpha$ values, all stellar types increase with increasing $\alpha$. Perhaps the only exception would be the number of ONeWD companions, which show the highest value for $\alpha = 0.2$. The ratio of this number against the number of COWD companions might shed some light on the CE process.

Other interesting features are the ratio of initially primaries to initially secondaries that become sdB candidates, and the progenitor stellar type. The ratio shows an increasing value with $\alpha$: low $\alpha$ favours initial secondaries becoming candidates (the opposite is true for high $\alpha$), potentially because only those systems where the initially primary star quickly evolves to the WD stage can remove the envelope of the secondary during a CE without merging; while the progenitor stellar type shows a preference for FGB stars with HG (CHeB) as second option for low (high) $\alpha$.

\subsubsection{Metallicity changes}\label{sec:mettt}

\begin{figure*}[h]
\centering
\includegraphics[width=1\linewidth]{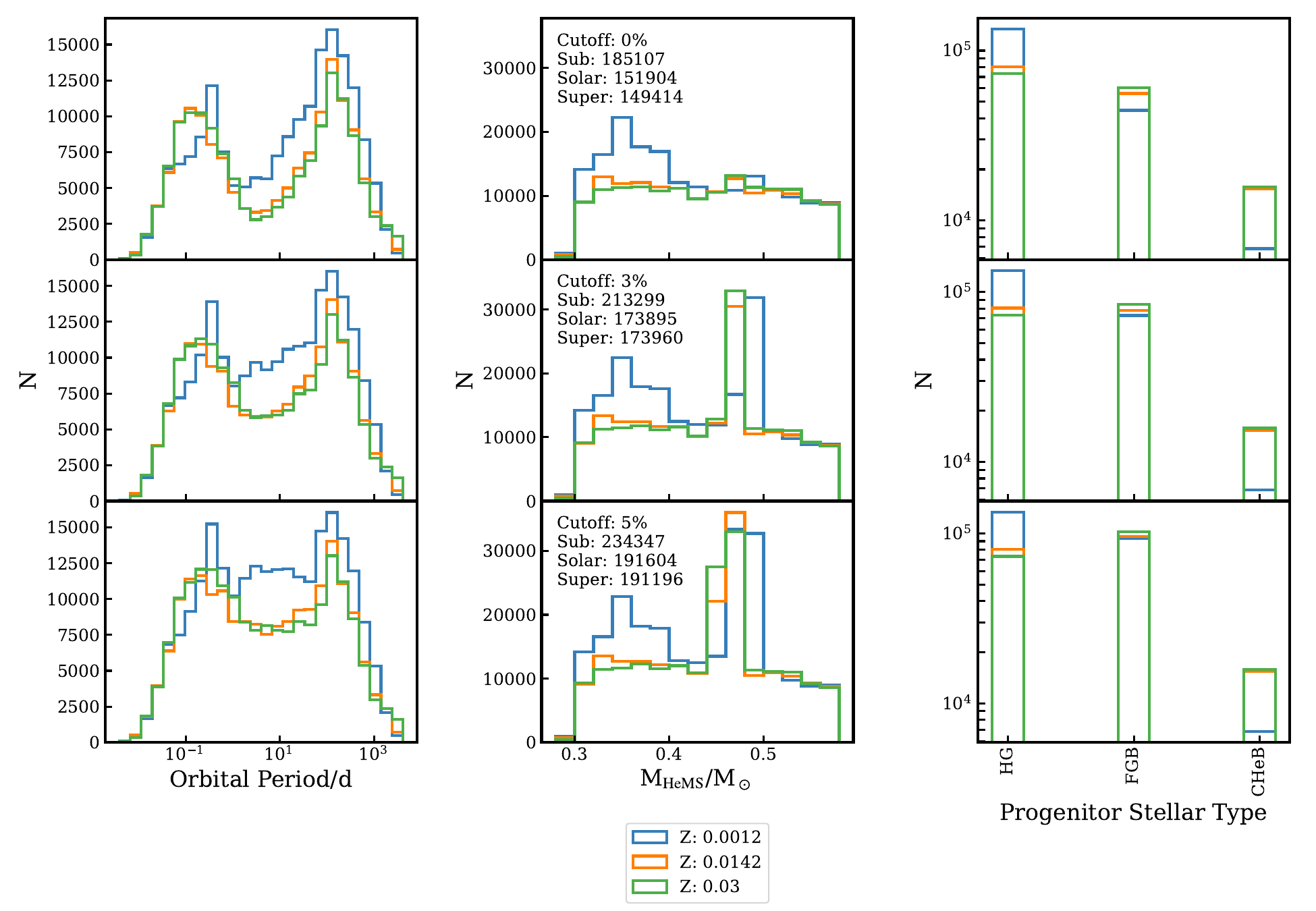}
\caption{From left to right: changes in orbital period, mass at HeMS and progenitor stellar type for our sdB candidates. The cutoff for stars that ignite helium in the core increases from top to bottom as a percent difference: 0\%, 3\% and 5\%. This can be understood as follows: the top panel shows candidates who have a mass equal or higher than the helium core mass expected at the tip of the RGB, while subsequent rows show the impact of considering HeWDs within 3\% and 5\% of this expected core mass as sdB progenitors. Different colors show different metallicities as indicated by the legend, with blue being sub-solar ($\rm{Z} = 0.0012$), orange solar ($\rm{Z} = 0.0142$), and green super-solar ($\rm{Z} = 0.03$). See sections \ref{sec:mettt} and \ref{sec:ignition_below} for details.}
\label{fig:metallicities}
\end{figure*}

\begin{figure}[h]
\centering
\includegraphics[width=1\linewidth]{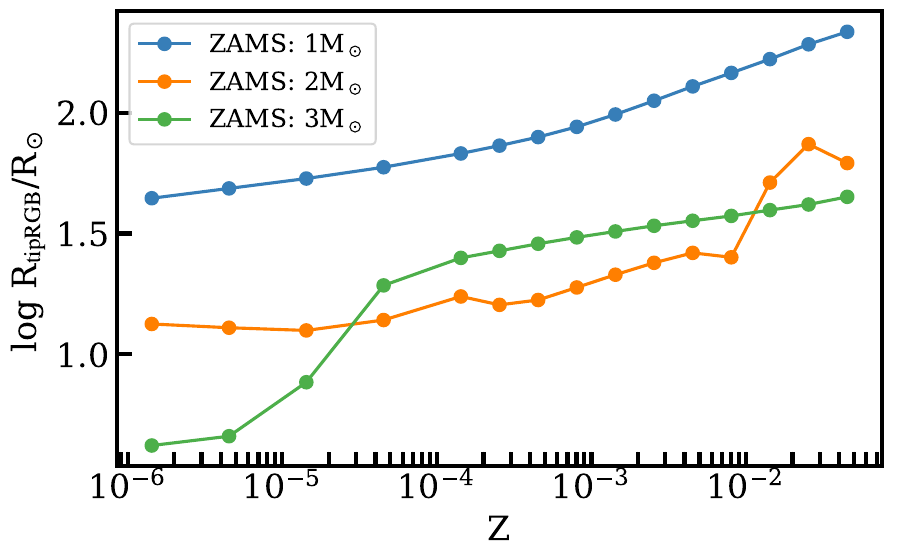}
\caption{Dependence of the stellar radius at the tip of the RGB on metallicity, as shown by the evolutionary tracks from the \textsc{mist} grids \citep{Choi2016}. We have chosen tracks that were computed and not interpolated, extracted metallicity values from the stored \texttt{Zinit} property, and interpreted the radius at tip of the RGB as the last (age-wise) radius value for which the corresponding time step is classified as having a \texttt{phase} equal to 2 (values equal to 2 consider both the sub-giant and red giant phase, see Choi et al. \citeyear{{Choi2016}} for details).}
\label{fig:rad_evo_mist}
\end{figure}

As mentioned in section \ref{sec:z}, metallicity is expected to play a role in a few different settings. The helium core mass and the size of the star at the tip of the RGB should both depend on the chosen metallicity value and, similarly, metallicity plays a role in defining the MHeF critical mass limit for degenerate or smooth helium ignition. Indeed, some of these effects can be seen in results for detailed models, such as what can be found in the \textsc{mist} stellar evolutionary tracks \citep{Choi2016}, explicitly shown in Fig. \ref{fig:rad_evo_mist}. 

As an overview of the effect of metallicity on our populations, we can say that there is a decrease in the total number of candidates with increasing metallicity (the number of candidates per metallicity are shown in the left-hand side column of Fig. \ref{fig:metallicities}). Taking the analysis one step further, within the top panels of Fig. \ref{fig:metallicities} we compare grids that share the same configuration but vary their metallicity value, and we can see that most of the stars that no longer become sdB candidates at higher metallicities are those that evolve from HG progenitors. These stars usually result in sdBs with masses below $\sim0.43 {\rm M}_\odot$, meaning that they must have evolved from progenitors with masses close to the MHeF limit (Fig. \ref{fig:core_range_and_mass}). It can also be seen that the \textit{canonical mass} peak, around $\sim0.47{\rm M}_\odot$, is displaced towards higher masses at lower metallicities, while the peak located at short orbital periods ($\lesssim 1$ day) is slightly displaced towards longer periods. Within the same period distribution, a gap forms between 1 to 10 days, which becomes more pronounced for higher metallicity values. The gap is less evident for larger values of the threshold for helium ignition (section \ref{sec:ignition_below}), but the trend with metallicity remains. We note that beyond slight differences in the total number of sdB candidates, there are not many remarkable differences between our supersolar and solar metallicity population, most likely due to the rather small 2.1 multiplicative factor required to scale our solar value to the supersolar one. This contrasts with our subsolar metallicity value, where the required factor is 0.08.

\subsubsection{Mass lost from the system}\label{sec:ml_sys}

Equation \ref{eq:axis_change} shows the interplay between mass being transferred from the donor to the accretor and what fraction is kept in the latter. In cases where the mass is not completely retained, the equation also takes into account the location at which this mass is being lost from the system and establishes the amount of specific angular momentum lost. These properties are encoded by $\beta$ and $\gamma$'s definition, as shown in equations \ref{eq:mteff} and \ref{eq:gamma}, respectively. However, it is not straightforward to predict the final effect of equation \ref{eq:axis_change} for the orbit, due to its non-linear dependence on the $\beta$ and $\gamma$ parameters, the mass of each component, and the mass loss rate of the donor. Instead, we can perform a simple analysis of Equation \ref{eq:axis_change} to estimate whether the rate of change of the semi-major axis will be negative or positive (the orbital separation shrinks or grows) after a mass transfer episode. First, the factor outside the square brackets on the right-hand side of Equation \ref{eq:axis_change} is always $\geq 0$, considering the minus sign and that mass is being lost from the donor (its mass change rate is negative). Then, we define a mass ratio $q = {\rm M}_{\rm a}/{\rm M}_{\rm d}$ and analyze the sign of the expression within the square brackets in Equation \ref{eq:axis_change}, referenced as $\phi$ hereafter. We can see that:

\begin{align}
    \phi = 1-\frac{\beta}{q}-\frac{\left(1-\beta\right)\left(\gamma+0.5\right)}{1+q}.
    \label{eq:phi}
\end{align}

\begin{figure*}[h]
\centering
\includegraphics[width=1\linewidth]{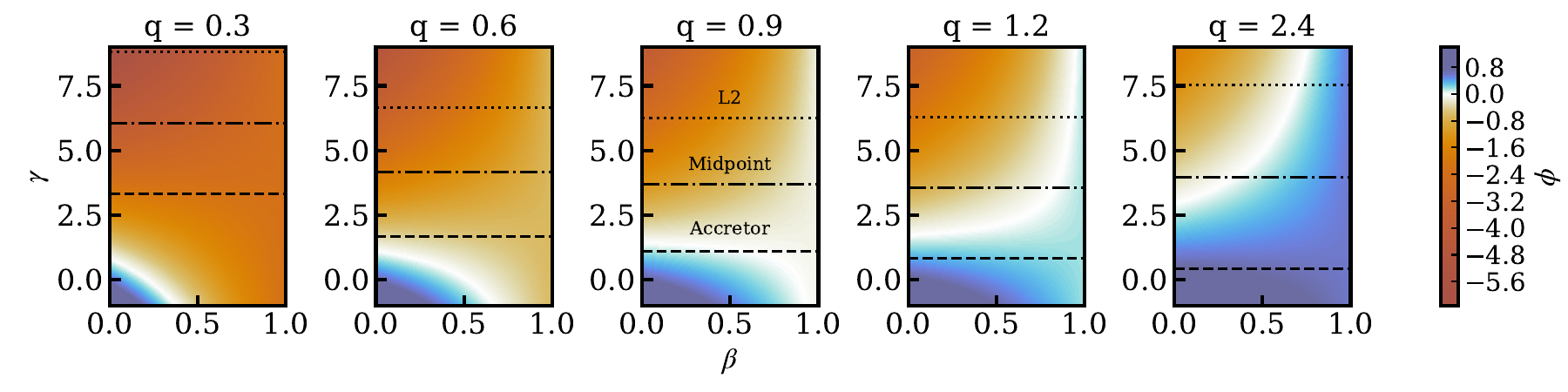}
\caption{Contour plots of $\phi$ as a function of $\beta$ and $\gamma$, for a given mass ratio (q). Representative values of $\gamma$ are shown for reference, depending on where mass is being lost from the system: the dotted line corresponds to L2, the dashed line to the position of the accretor, and the dash-dotted line to the midpoint between the previous two. The more positive (bluer) $\phi$ is, the more the orbital separation grows. The opposite is true for negative (orange) values. For details, see section \ref{sec:ml_sys}.}
\label{fig:gamma_beta}
\end{figure*}

A visualization that helps to understand the values that $\phi$ can take is shown in Fig. \ref{fig:gamma_beta}, where it can be seen that for most of the configurations being shown $\phi$ is negative and, as a consequence, the orbital separation should shrink. This trend is reversed as we progress into the mass ratio reversal regime ($q$ becomes greater than 1). It should be noted that for this analysis we have assumed that $\gamma$ is equal to the upper limit shown in \citet{Willcox2023} when MLF is equal to 1. Accordingly, $\gamma$ takes the value $1/q$ when mass is lost from the position of the accretor, implying that MLF is equal to 0. MLF equal to 0.5 follows from the previous two definitions, as the intermediate value.

\begin{figure*}[h]
\centering
\includegraphics[width=1\linewidth]{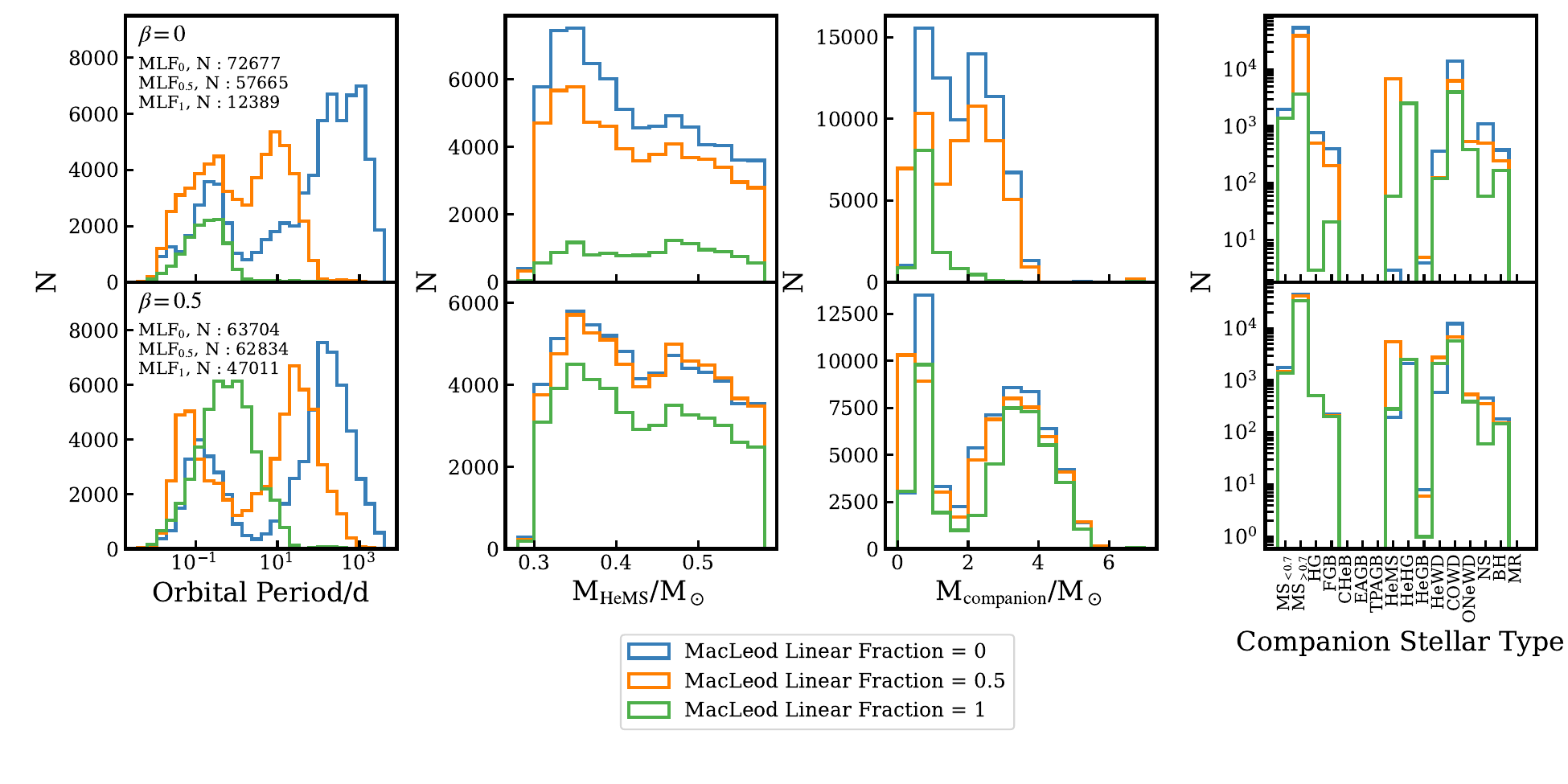}
\caption{Graphical analysis of the impact of changing the MLF at a constant $\beta$ value. From top to bottom: $\beta = 0$ and $\beta = 0.5$. From left to right: Orbital period, mass of the candidate at the start of the HeMS, companion's mass at the start of the HeMS, and companion stellar type at the same evolutionary stage. Different colours indicate different values for the MLF, with the number of candidates per configuration being specified within the text of the left-hand side panels.}
\label{fig:gamma_changes}
\end{figure*}

\begin{figure*}[h]
\centering
\includegraphics[width=1\linewidth]{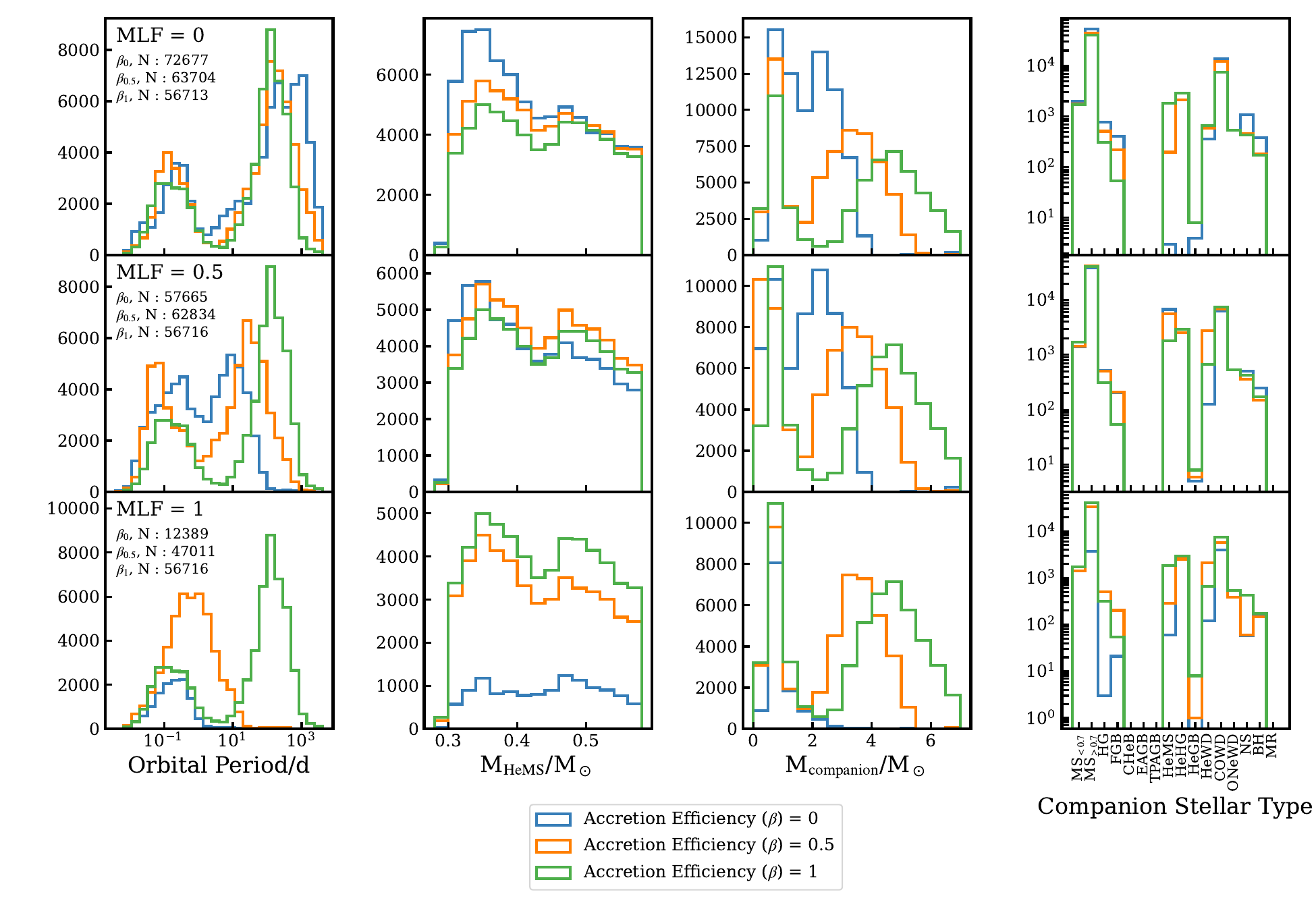}
\caption{Similar to Fig. \ref{fig:gamma_changes}, graphical analysis of the impact of changing the \textit{Accretion Efficiency} ($\beta$) at a constant MLF value. From top to bottom: ${\rm MLF} = 0$, ${\rm MLF} = 0.5$ and ${\rm MLF} = 1$. From left to right: Orbital period, mass of the candidate at the ZAHeMS, companion mass at the ZAHeMS, and companion stellar type at the same evolutionary stage. Different colours indicate different values for $\beta$, with the number of candidates per configuration being specified within the text inside the left-hand side panels.}
\label{fig:beta_changes}
\end{figure*}

To analyze the impact of changing $\beta$ and $\gamma$ in our different \textsc{compas} populations, we present Figs. \ref{fig:gamma_changes} and \ref{fig:beta_changes}, which depict how some physical properties change when keeping $\beta$ constant and varying $\gamma$, and vice versa. Before proceeding with the analysis, however, it must be noted that the mass accretion efficiency parameter in \textsc{compas} only affects stable mass transfer episodes. Unstable mass transfer always results in no mass accreted in \textsc{compas}\footnote{However, it is possible to change this behavior for NS/BH accretors through the \texttt{--common-envelope-mass-accretion-prescription} option in \textsc{compas}, which is set to \texttt{ZERO} (disabled) by default.} \citep{Riley2022}.

In the case of keeping $\beta$ equal to zero and varying $\gamma$ (top panels in Fig. \ref{fig:gamma_changes}), we note that the number of candidates drops as we increase the latter (in the form of MLF), which can be understood as fewer systems surviving in closer orbits due to the impact of $\phi$ continuously decreasing. This can be seen in Fig. \ref{fig:gamma_beta} where $\phi$ becomes increasingly negative for $q \lesssim 0.9$ as MLF ($\gamma$) increases. A similar trend is present for higher values of $q$, though in these cases the lowest MLF values start at $\phi \gtrsim 0$ and transition into negative values as MLF grows. Overall, this results in many systems from the top panel in Fig. \ref{fig:gamma_changes} either merging or following different evolutionary paths as a consequence of the increase on MLF: the long period peak at about 1000 days when MLF equals 0 is shifted towards 10 days when MLF increases by 0.5, and the new peak is not as tall. This effect is further strengthened by losing mass from L2 instead (MLF$ = 1$), which results in a single peak below 1 day. Regarding the mass distribution, it seems to have a common pattern with two peaks, no matter the value of $\gamma$, though it can be noted that the low-mass peak is comparatively higher than the peak around the canonical ($\sim 0.47$M$_\odot$) mass for the MLF$ < 1$ scenarios. This difference is smaller for increasing $\gamma$ values, and shows a reversal in peak prominence when MLF$ = 1$, a trend that can be linked to a preference for progenitors with masses around the MHeF limit when MLF takes values lower than 1. As for the characteristics of the companions, their mass distributions peak at $\sim 1$M$_\odot$ independent of the MLF choice, though a secondary peak can be observed at about 2M$_\odot$. This additional peak is of similar relevance to the low mass one when MLF$=0.5$, but completely vanishes for MLF$=1$. Overall, both peaks decrease with increasing MLF value. These changes seem to be driven by a drop in the number of companions in stages of evolution earlier than the FGB.

When looking at a constant 0.5 value for $\beta$ (bottom panels in Fig. \ref{fig:gamma_changes}), a similar analysis can be made. The number of candidates also decreases as we increase $\gamma$, but the differences become smaller and are almost negligible between MLF equal to 0 and 0.5. This can be seen in Fig. \ref{fig:gamma_beta}, as the colour corresponding to $\phi$ values does not change as much as in the case of $\beta = 0$ (for increasing $\gamma$). By looking at the significance of the long period and short period peak at each MLF value shown in the period distribution, it can be seen that smaller MLF values favour longer periods since $\phi$ is closer to 0 as $q$ grows, and it can take positive values once $q \gtrsim 1$. However, the displacement of each peak does not follow a clear dependence on MLF, unless the single peak present for MLF $= 1$ corresponds to the long period peak of the other two MLF values, in which case it would have been shifted enough to merge with the short period peak, implying closer orbits for larger MLF values overall. The mass distribution shows a similar pattern to that observed when $\beta$ is equal to 0, though the relative significance of the two peaks (at about $0.35 M_\odot$ and canonical mass) is rather constant. Additionally, the number of systems for MLF$= 1$ is much higher than in the $\beta = 0$ scenario. Finally, the companions show two peaks in their mass distribution, one in the 0 -- 1 M$_\odot$ mass range and another at about 3.5 M$_\odot$. Both decrease their prominence with increasing MLF values.

The case of constant $\beta = 1$ is not analyzed as changes in $\gamma$ would not impact the distributions. This can be understood through either Equation \ref{eq:axis_change} or \ref{eq:phi}, since they show that setting full accretion efficiency implies that there would be no mass being lost from the system that could carry a fraction of angular momentum away during a mass transfer event.

We can also probe the effects of different mass retention efficiencies ($\beta$) by keeping MLF constant in a similar fashion to what was done before, though the analysis this time requires considering Fig. \ref{fig:beta_changes} instead.

When losing mass from the accretor's position (MLF = 0), there is a clear preference for periods longer than $\sim 100$ days, though we notice that there is a slight dependence of the number of candidates on the value of $\beta$, as they decrease with increasing $\beta$. This results in period and mass distributions without striking differences between configurations. The biggest differences between $\beta = 0$ and the other configurations are the number of candidates in both intermediate ($\sim 1 - 10$ days) and long ($\sim 1000$ days) periods, or low-mass sdBs that peak at $\sim 0.35 {\rm M}_\odot$ in the mass distribution. Another expected characteristic is that a larger $\beta$ value results in more massive companions due to larger masses being retained, as can be observed in the displacement of the secondary peak towards higher mass values in the companion mass distribution with increasing $\beta$.

Instead, if every system is set to lose mass from an intermediate location between the accretor and L2 (MLF = 0.5), the changes on the number of candidates between the different accretion efficiency values are even smaller than in the MLF$=0$ case, and there is no clear preference for long-period systems anymore. Despite this, the differences between different choices of $\beta$ are clearer in the distributions. We can see two peaks that have a similar significance when $\beta = 0$, and progressively show a more relevant long period peak with increasing $\beta$. Along this trend, the long period peak is also shifted to longer periods, from an initial peak about 10 days (when there is no accretion), up to $\gtrsim 100$ days when there is full accretion. This can be associated to less angular momentum being carried away by mass lost from the system during mass transfer events for higher values of $\beta$ and, as a consequence, orbits not shrinking as much. However, the HeMS mass distribution seems to have the same set of features, independent of the choice of $\beta$. Of course, the frequency per bin changes depending on the total number of candidates, but the presence of two peaks ($\sim 0.35 M_\odot$ and around canonical mass)  remains unaltered. There seems to be a slight dependence on relative peak prominence with accretion efficiency, though: higher $\beta$ values create peaks of increasingly similar height. The companion mass distribution is rather similar to the MLF$ = 0$ case, as we can notice that higher accretion efficiency values result in more massive companions (shown by the displacement of the secondary peak towards higher masses).

Finally, by setting MLF = 1, we find the biggest changes in the number of candidates per accretion efficiency configuration, as $\beta = 0$ yields around 20\% of the total number of candidates that can be obtained by setting $\beta = 1$ instead. It is worth noting that an interesting behavior can be observed in the period distribution: if we consider the $\beta = 0$ case as the base configuration, then increasing the accretion efficiency up to a 0.5 value creates more short and intermediate period systems (up to $\sim 10$ days) and keeps the distribution as a single peaked one. This changes in the full accretion efficiency case ($\beta = 1$), since an additional long period peak can be found, but the short period peak is almost equal to the base scenario. This is most likely another form of the patterns seen in the bottom panel of Fig. \ref{fig:gamma_changes} and discussed earlier in the text, though in this case the long period peak at full accretion is displaced to shorter periods at $\beta = 0.5$, giving the impression of  a single peak. Many systems composing this peak are unable to generate sdB candidates when there is no accretion, hence the much smaller (and single) peak in this case. The mass distribution of the candidates follows the same pattern as in the previous MLF configurations, except for the case without accretion which breaks the trend and shows a much flatter distribution. Interestingly, the mass distribution of the companions does not show a relevant number of massive $\gtrsim 2{\rm M}_\odot$ companions for the case without accretion.

\subsubsection{Mass Transfer Stability Prescription}

\begin{figure*}[h]
\centering
\includegraphics[width=1\linewidth]{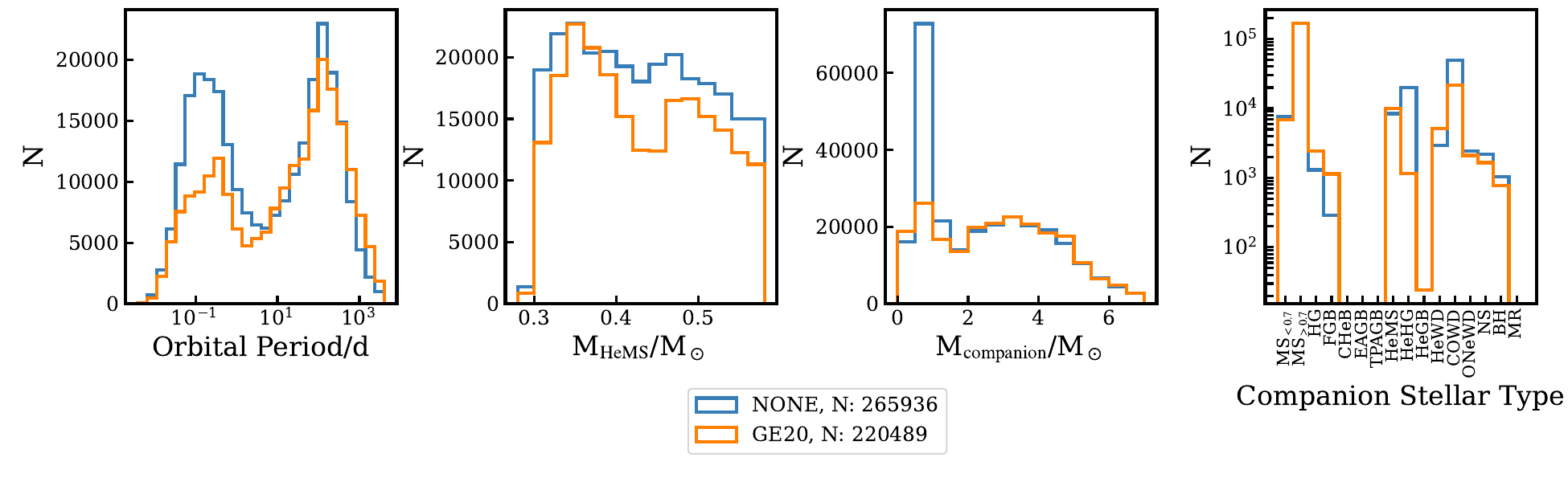}
\caption{Similar to Figs. \ref{fig:gamma_changes} and \ref{fig:beta_changes}, each panel from left to right corresponds to orbital period, mass of the candidate at ZAHeMS, mass of the companion at ZAHeMS, and the companion's stellar type at the same stage. In this case, different critical mass ratio prescriptions are being tested: in blue, the $\zeta$ prescription labeled as \texttt{NONE} due to how it is implemented in \textsc{compas} and, in orange, the \citet{Ge2020} prescription (see section \ref{sec:qcrits} for details).}
\label{fig:qcrit_changes}
\end{figure*}

As the name indicates, the choice for mass transfer stability prescription determines whether a star experiences a stable or unstable mass transfer episode after filling its Roche lobe. This, in turn, affects the number of sdB candidates and their characteristics, as can be seen in Fig. \ref{fig:qcrit_changes}. In rough numbers, the $\zeta$ prescription \citep[e.g.,][labeled as \texttt{NONE} in Fig. \ref{fig:qcrit_changes}]{Soberman1997} produces around 1.2 times more candidates than the \texttt{GE20} \citep{Ge2020} prescription, though these differences are concentrated around channels involving CE events: the \texttt{NONE} prescription yields 1.6 and 2.5 times more candidates than the \texttt{GE20} one for 1 and 2 CE episodes pre-candidate formation, respectively. Moreover, if analyzed in detail, the overall number of candidates is not the only difference. First, \texttt{GE20} seems to prefer long period systems. This is not the case for the $\zeta$ prescription, as the preference for long orbital periods is not so evident and the relative significance of both long and short period peaks is similar. Similarly, the mass distribution of sdB candidates seems to decay almost linearly in the case of the $\zeta$ prescription, but has two clear peaks if the \texttt{GE20} distribution is considered instead.

The properties of the companions also depend on the chosen mass transfer stability prescription. It is clear that the \texttt{GE20} prescription yields less systems overall, linked to the $\zeta$ prescription allowing the existence of a sharp and tall peak at $\sim 0.7M_\odot$ companion masses. This also indicates differences in the stellar types of the companions, confirmed by the right-hand side panel of Fig. \ref{fig:qcrit_changes}: the $\zeta$ prescription generates a higher number of COWDs as well as HeHG companions.

\subsubsection{Ignition Below The Expected Core Mass}\label{sec:ignition_below}

As stated in section \ref{sec:corehe}, the analysis of our sdB candidates requires considering the possibility of stars igniting helium burning when their core mass is below the expected value near the tip of the RGB. We have tackled this issue by collecting stars originally classified as helium white dwarfs (HeWDs), and used their mass values as input for our new HeMS with hydrogen-rich shells prescription, which returns evolutionary tracks as if they were HeMS stars instead. We only consider HeWDs that are within the mass cutoff of either 3 or 5\% that we set based on results from section \ref{sec:corehe} and previous studies \citep{Han2002,ArancibiaRojas2024}. While it is possible to follow a similar approach to \citet{Zorotovic2013}  and use the the actual values found for minimum helium ignition mass in either \citet{Han2002} or \citet{ArancibiaRojas2024} , we note that both studies present a $\sim 5$\% threshold for stars igniting helium in a flash, independent of the chosen metallicity value, hence our choice of 5\% for one of the tested configurations. The case of more massive progenitors that ignite helium smoothly is slightly more complicated, as according to \citet{ArancibiaRojas2024} there is a wider range of masses for ignition (their tables B1 and B2). On top of that, the width of this range also depends on the ZAMS mass of the model: there is a linear increase from 5\% at ZAMS masses corresponding to the most massive progenitors that ignite helium in a flash, up to $\sim 30$\% at about 3 M$_\odot$. This value then decreases at a slower rate (it is $\sim 15$\% at 6M$_\odot$), but the trend is nonetheless completely different from the relatively flat 5\% value for models igniting helium in a flash. For simplicity purposes, and noting that the realization probability of more massive progenitors is much lower due to the initial mass function (IMF), we decide to use flat threshold values. The impact of these different threshold values on our population can be seen in Fig. \ref{fig:metallicities}, where a considerable spike in the number of candidates around canonical mass can be observed for cutoff values $> 0$\%, independent of the chosen value for metallicity (histograms of different colours). As expected, increasing the percentage for the mass cutoff increases the total amount of candidates, though interestingly this increase is focused on stars with masses around the canonical sdB mass value ($\sim 0.47\rm{M}_\odot$) and intermediate orbital periods between $\sim 1 - 10$ days only. It can be inferred that the observational sdB sample and its mass distribution would be directly impacted by the real value of the mass cutoff, though additional constraints such as the delay times associated to the formation of these objects play a non-negligible role. Additionally, it is possible to see that the peak in the mass distribution is shifted depending on the chosen value for metallicity, as a consequence of the arguments discussed in section \ref{sec:mettt}.

We remark that this is a simple approach to the underlying question of the lower limit for helium ignition and it represents a first order estimate. As previously mentioned, the results from \citet{ArancibiaRojas2024} show that stars with ZAMS mass above the MHeF mass limit require a broken linear function with different slopes for mass values above/below $\sim 3$M$_\odot$, though in a Galaxy-like sampling these systems would not be as common due to the IMF.

\subsubsection{Hydrogen-rich Shell Mass}

\begin{figure}[h]
\centering
\includegraphics[width=1\linewidth]{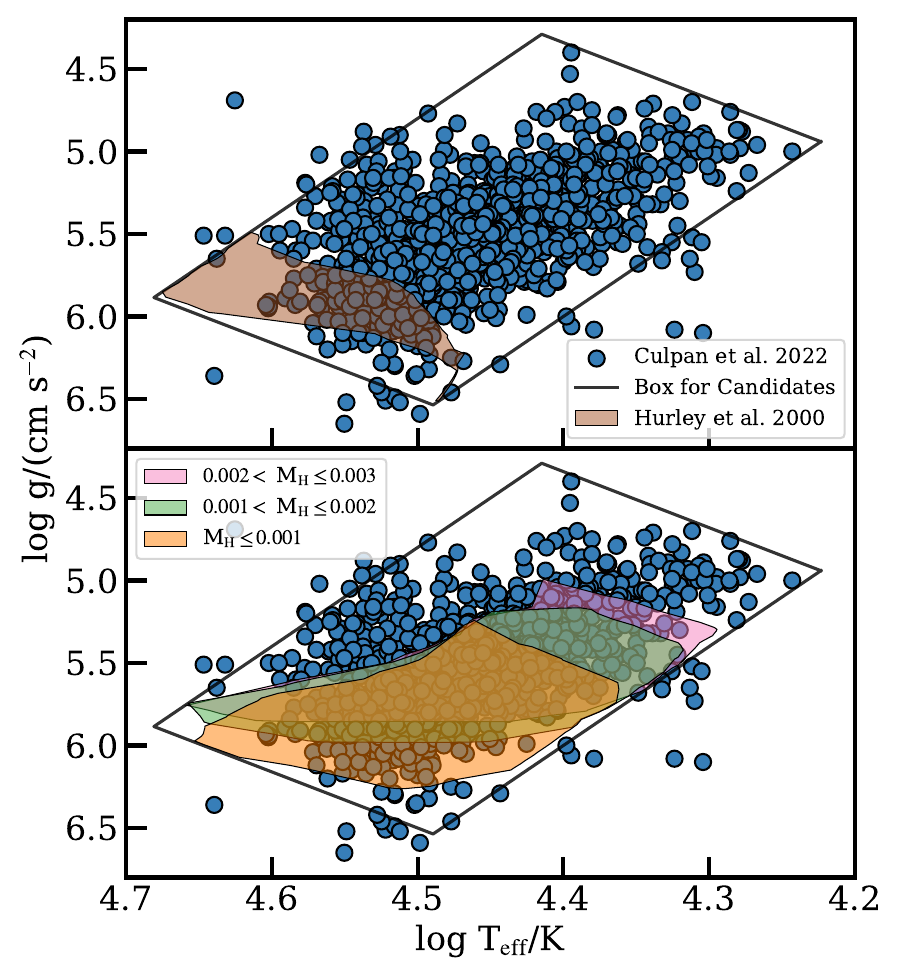}
\caption{Comparison between the physical properties of the \textsc{compas} sample when no hydrogen-rich outer layers are considered (top panel) and when they are included by using the prescription presented in section \ref{sec:h_prescription} (bottom). This last element is further explored by presenting the area covered by the \textsc{compas} sdB candidates depending on their M$_{\rm H}$ value, which corresponds to the hydrogen-rich outer envelope's mass (in Solar mass units). Scatter points (in blue) are taken from the \citet{Culpan2022} observational sample and shown for reference.}
\label{fig:kiel_theo}
\end{figure}

The definition of HeMS stars in the \cite{Hurley2000} scheme adopted by \textsc{compas} implies a complete absence of any hydrogen-rich envelope, due to the prescription for their evolution being based on detailed models of ZAHeMS stars ($Y = 0.98$ and $Z = 0.02$ composition). Although this is a useful simplification in the context of rapid BPS codes, it yields estimates for effective temperature and surface gravity that are different from observations (e.g., Fig. \ref{fig:kiel_theo}, upper panel). Indeed, the absence of any external hydrogen-rich shells makes the candidates from simulations cluster at high temperature and surface gravity, unlike the spread of the observed distribution. The inclusion of hydrogen-rich shells through the prescription based on \cite{Bauer2021} and developed in section \ref{sec:h_prescription} seems to properly address the issue and yields results that improve the agreement with observables, as can be seen in the bottom panel of Fig. \ref{fig:kiel_theo}. As one might intuitively expect, the more massive the envelope is, the less compact and colder a star can become. This effect can be seen through the location and coverage of each patch within the figure and, although the coverage is not perfect, we must keep in mind that the current state of the prescription is limited in mass range as well as envelope size (see section \ref{sec:h_prescription}). This lack of coverage at lower surface gravity and temperature in Fig. \ref{fig:kiel_theo} suggests that higher mass envelopes are needed, and will be addressed in future work.

It is important to note that the impact on the total number of sdB candidates is almost negligible, as the inclusion of hydrogen-rich shells only increase the number of sdB candidates by 1\%. Also, these hydrogen-rich envelopes do not affect the orbital evolution of the systems. After the onset of a mass transfer episode, the external low-density hydrogen-rich shell would be slowly transferred during a few tens of mega-years, while the orbit shrinks due to gravitational wave radiation \citep[see][section 3.1]{Bauer2021}.

Finally, an important question left to answer is related to the mass distribution of these hydrogen-rich envelopes. The different sdB formation channels involve different interactions and progenitors at different evolutionary stages, which would potentially impact the preferred size of the hydrogen-rich layers (as previously noted in H02). We made no special assumptions and limited ourselves to randomly sampling masses in the range 0 to $3\times10^{-3} {\rm M}_\odot$, which corresponds to the values covered by the \citet{Bauer2021} models. 

\subsection{Formation Channels}\label{sec:channels}

\begin{figure*}[h]
\centering
\includegraphics[width=1\linewidth]{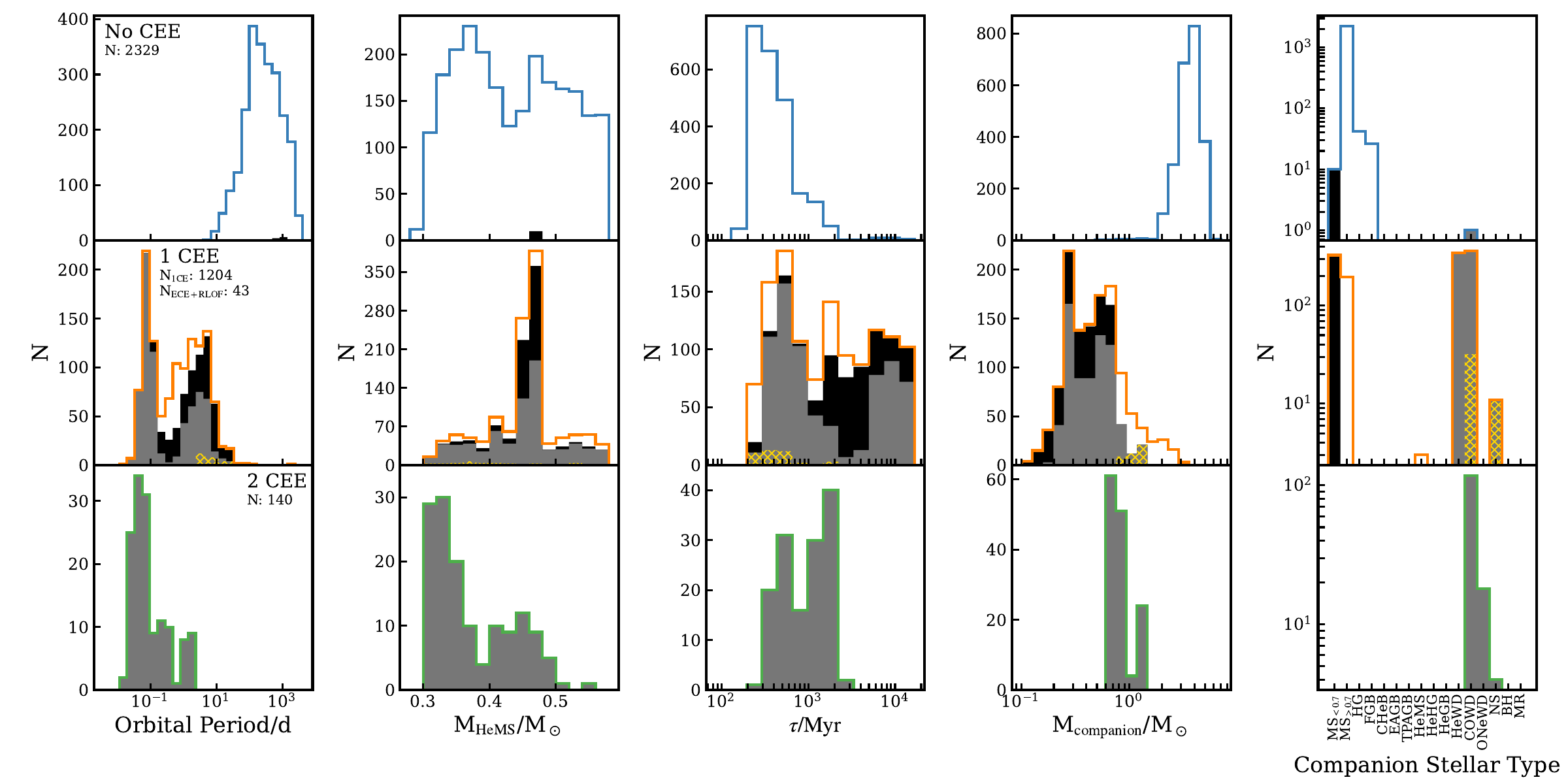}
\caption{Different sdB formation channels found in our \textsc{compas} chosen model. From left to right, columns contain the orbital period, the stellar mass on the ZAHeMS, total time it takes to form an sdB candidate, mass distribution of the companions, and the stellar type of the companion when the candidate starts the HeMS. From top to bottom, we show an increasing number of CE events that led to the candidate's formation: No CE episodes (stable mass transfer only), one CE episode and two common envelope episodes. To highlight different companion types, we have used different shades: the lightest shade corresponds to Stellar Type $>9$ (white dwarfs, neutron stars, or black holes), while the darkest shade corresponds to Stellar Type $= 0$ (MS companions with masses $\leq 0.7 M_\odot$). A white colour indicates types in-between (check table \ref{tab:sttypes} for definitions, though most of these companions correspond to MS stars with masses $> 0.7 M_\odot$), and the sum of the contribution of each one of these 3 colours corresponds to the total amount of systems in a given bin. A hatched yellow area has been used to highlight candidates evolving through the ECE+RLOF channel, defined in section \ref{sec:onece}. Note that in each panel the y-axis has a different range.}
\label{fig:set}
\end{figure*}

Considering the analysis in section \ref{sec:par_var}, we can see that all the studied parameters impact the final yields of our population and therefore a more in-depth analysis of sdB formation channels should only be performed after selecting a specific configuration. For this section, we chose a model based on what we consider a \textit{neutral} configuration: solar metallicity, $\alpha_{\rm CE} = 1$, mass being lost from the position of the accretor during mass transfer, semi-efficient mass retention ($\beta = 0.5$), and critical mass ratios computed through the \texttt{GE20} prescription. Additionally, we consider the hydrogen-rich shell prescription a necessary component, as it impacts the number of HeMS stars crossing our sdB candidate selection area in the Kiel diagram.

With this configuration, we find sdB formation channels similar to those described in the literature (e.g., H02): different combinations of episodes of stable mass transfer and CE episodes. Some relevant properties of these channels are shown in Fig. \ref{fig:set}, which we analyse within the following subsections.

\subsubsection{Stable Mass Transfer Only}

In this case, all our systems containing sdB candidates correspond to rather long period systems, between 10 to 3000 days. The logarithmic period distribution peaks at about 120 days, and is slightly skewed towards longer periods. Most progenitors (about 70\%) belong to the HG stage, followed by those on the FGB. Our results show that progenitors with ZAMS masses above the MHeF value are more likely to evolve into an sdB from an HG progenitor, in this case.

When it comes to the mass distribution of the sdB candidates, it is possible to distinguish two clear peaks. The first one is centered at around $0.37M_\odot$, while the second one is located at $0.47M_\odot$. The low mass peak is closely related to the aforementioned HG progenitors, while the second peak is close to the canonical mass. Finally, we note that most of these systems are formed within the initial $\sim1$ Gyr of their lifetime ($\tau$ column in Fig. \ref{fig:set}), with a peak at $\sim 250$ Myr in logarithmic space, which implies that they should be more abundant in young components of the Galaxy. However, their real relevance within the current day population can only be estimated by studying the star formation history of the Galaxy. 

\subsubsection{One Common Envelope Episode}\label{sec:onece}

For this channel, we find two different possible sub-channels: sdB candidate formation right after the CE (1CE), or formation after stable mass transfer in systems that also experienced an early CE episode (ECE+RLOF). Since the interaction that led to the formation of the sdB candidate is different, the orbital properties of such systems are different as well. This can be seen in the corresponding panels of Fig. \ref{fig:set}, where the ECE+RLOF channel is shown as a yellow hatched area, a subset of all the systems that experienced a single CE episode pre-sdB candidate formation. This subset is restricted to periods longer than about 2 days and, although it represents only a small fraction of the total number of systems, its relevance comes from the fact that most of the sdB+NS systems follow this formation channel. For reference, the time difference between a CE episode happening first and then a stable mass transfer being triggered in sub-channel ECE+RLOF is usually a few hundreds of mega years. Though the specific characteristics and relevance of this sub-channel depend on parameters such as the chosen value for accretion efficiency (table \ref{tab:params}), it consistently appears throughout our populations.

Candidates formed through sub-channel 1CE peak at about $0.47 M_\odot$ in the mass distribution, as most of the candidates within this peak come from sources initially classified as HeWDs, but considered as sdB candidates due to how close they are (5\%) to the expected core mass at the tip of the RGB. The situation is different for candidates from sub-channel ECE+RLOF, as the they peak at about 0.37 M$_\odot$, with a few candidates below the peak and the rest covering the whole range up to $\sim0.58 M_\odot$. 

As previously mentioned, an important fraction of the companions from sub-channel ECE+RLOF are NS, with the remaining companions belonging to the COWD class. In the case of 1CE sub-channel, companions are predominantly MS stars and WDs.

When looking at the $\tau$ distribution, sub-channel 1CE can generally produce sdB-candidates at almost any given age after $\sim 200$ Myr, though there are three spikes of sdB formation: one at the initial bin ($\sim 200$ Myr), another one around 1,800 Myr (the most prominent peak), and one above 6,000 Myr. On the other hand, the birth of sdBs from sub-channel ECE+RLOF is mostly concentrated between 200 -- 600 Myr, peaking at about 400 Myr.

\subsubsection{Two Common Envelope Episodes}

This channel occurs when the initially primary star overflows its Roche lobe at an earlier time than the sdB candidate formation. This initial event leads to the removal of the primary star's outer layers, triggering a first episode of unstable mass transfer (CE). Its outcome, however, is not the birth of the sdB candidate. Only when the initially secondary star fills its Roche lobe and the system undergoes a new CE episode, this time leading to the birth of an sdB candidate, we consider it as a member of this formation channel. In our chosen model, we get a total of 140 candidates from this channel, compared to 2329 from the stable mass transfer channel and 1247 in which only one CE episode is found in the system's evolution (pre-HeMS). That means that this channel accounts for $\sim4$\% of the sdB candidates within our chosen model (without considering potential mergers), showing that it is much less common than the other two when analyzing the total numbers. Notably, these systems are mostly formed in the 300-2000 Myr age range and peak at about 1800 Myr.

A clear preference for low mass HeMS stars can be seen in Fig. \ref{fig:set}, with a peak around $0.33 M_\odot$ and a sharp decline after the peak. A second peak seems to be present at $\sim 0.45 M_\odot$, though it might be cause by the low number of members of this channel. The progenitors of these candidates are all FGB stars with ZAMS masses higher than $1.5 M_\odot$ that peak around $2.3 M_\odot$, which is slightly lower than what can be found for other channels but in line with the progenitor star being the initial secondary (less massive) star. The masses of the companions at sdB-candidate formation are considerably low as well, most likely owing to the first CE episode where they should have lost a fraction of their mass during the envelope removal. Indeed, the companions are mainly COWDs with masses in the range 0.5 -- 1.5 M$_\odot$. 

As one would expect, the 2 CE events considerably shrink the orbital separation, and thus the period distribution peaks at about 0.04 days, with the shortest (largest) period being $\sim 0.01$ (2) days. 

\subsubsection{Mergers}

\begin{figure}[h]
\centering
\includegraphics[width=1\linewidth]{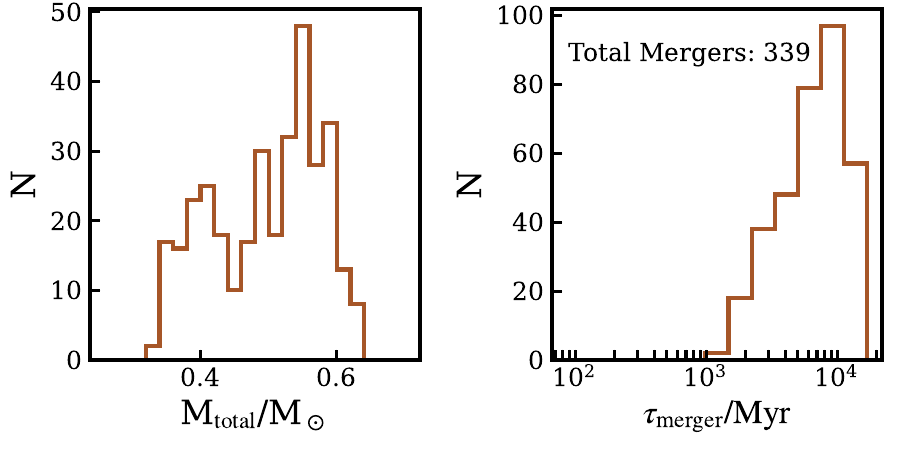}
\caption{Double HeWD mergers from our chosen model, selected as systems with the \texttt{Merger} flag triggered in \textsc{compas}. Only systems that merge within 13.5 Gyr have been considered. The left panel shows the expected mass (sum of the masses of the merging HeWDs), while the right panel depicts the time that system requires to merge (since birth), e.g. the merger delay time distribution.}
\label{fig:mergers}
\end{figure}

The current version of \textsc{compas} does not take into account evolution beyond the formation of double WD systems and, therefore, analysis made in this subsection has been done through a simplified post-processing pipeline for informative purposes only. 

First, we note that while there are ${\sim}$40 double HeWD systems in our chosen model, only 4 of these systems have merger times below 1 Gyr \citep[based on equation 5.10 from][]{Peters1964} and thus are worth considering as plausible sdB progenitors. All remaining systems would require a time longer 14 Gyr to merge, which is a striking result, considering the non-negligible percentage of sdBs formed through mergers in H03. However, in this \textit{conservative} estimate, we have not considered the consequences of excluding merging and touching stars (according to the definition in \textsc{compas}) from our initial sample of candidates. If we analyze these removed double HeWD systems, we get the results from Fig. \ref{fig:mergers}, where it can be seen that a total of 339 sdB candidates would be formed through the HeWD+HeWD merger channel. 

For the merger products, the mass distribution (computed as the sum of merging white dwarf masses) seems to be composed of two groups, one peaking at about $0.41 M_\odot$ and the other at $\sim 0.55 M_\odot$. For completeness purposes, we checked whether these two groups were present in cases where the accretion efficiency is 0 or 1, and found that the group peaking at low mass is strongly dependent on the choice of this value. The no-accretion case completely removes the low mass peak, while full accretion enhances it up to the point of making the distribution look symmetrical and single-peaked around $\sim 0.47 M_\odot$. However, the total time it takes a binary system from ZAMS to the merger event is strictly longer than 1 Gyr and up to 13.5 Gyr (the upper time limit we set in our simulations), with a peak at $\sim 10$ Gyr, no matter the accretion efficiency choice. As what was done for the systems in our chosen model, these estimates have been computed using the same \citet{Peters1964} equation plus the time it takes the system to form a pair of HeWDs in \textsc{compas}. 

\subsection{Comparison To Previous Studies}\label{sec:comp}

In line with similar work available in the literature, such as H02 and H03, we have found that channels leading to the formation of an sdB can be classified by the number and type of mass transfer events required to almost completely remove the outer hydrogen-rich layers of the progenitor. It must be noted that in section \ref{sec:channels} we only distinguish between the number of CEs and do not make the exact same analysis as the authors of H02 do, though we find general agreement between our results and the summary of sdB formation channels presented in H03.

First, we compare against their \textit{first CE ejection channel}, which is roughly equal to our previously defined 1CE sub-channel: when the binary system has experienced a single CE episode that leads to the sdB candidate formation. However, for a better comparison, we select only those sdB candidates that correspond to the initially most massive (primary) star of the system, and we also include stars that were classified as HeWDs by \textsc{compas} (but can be considered as candidates according to section \ref{sec:ignition_below}, setting a 5\% threshold). With these constraints in place we find general agreement with properties presented in H03: most of the progenitors are classified as FGB stars (the exception being a few CHeB stars), the masses of the candidates show a sharp peak at $\sim 0.46 M_\odot$, and the companions are in the MS stage. For stars born from progenitors with masses below MHeF, we find a similar minimum period, on the order of $10^{-1}$ days (twice the H03 value), and our maximum period is around 30 days,  which is a bit different from what is reported in H03 as they mention a maximum period $\geq 40$ days. We note that this maximum period value can be increased by, for example, changing our choice of critical mass ratio prescription from \texttt{GE20} to \texttt{NONE}, raising our maximum period to around 50 days. Additional small discrepancies are found for candidates born from progenitors with ZAMS masses higher than MHeF, considering that they are described as products of HG stars resulting in low HeMS masses, while we find instead that our candidates are all in the FGB and cover a wide (0.3 to $0.6 M_\odot$) mass range without a well defined peak. We should remark, however, that we do find these systems with more massive progenitors restricted to short periods ($\lesssim 1$ days), as pointed out by H03.

We then proceed to analyze the \textit{first stable RLOF} channel presented by H03, described as being composed of sdB candidates with MS companions in wide orbits that cover the 0.5 to 2000 days period range. There is agreement if we compare against sdB candidates formed after the initially primary star overflows its Roche lobe and loses its outer layers through stable mass transfer in our chosen model, particularly finding companions in the same evolutionary stage (MS) and a similar orbital period range. In our case, the periods cover the 5 to $\gtrsim 3000$ day range, though it should be noted that there is a constant decrease in the number of systems for periods greater than 500 days. H03 also mention that the short period systems are stars going through the start of the HG and, although we have not verified in which section of the HG our candidates lost most of their outer hydrogen-rich layers, we find that the short period end of our period distribution (5 to $\sim 500$ days, peaking at about 100 days) is also composed of HG stars. For reference, candidates born from FGB stars cover the 50 to 3000 days period range, while we find fewer CHeB progenitors in the 500 to 3000 day period range. Going back to the HG progenitor group, H03 comment on a wide mass distribution range, which is also observed in our sample, though our upper sdB mass is restricted by our initial cut at $0.58 M_\odot$.

The analysis of H03's \textit{second CE ejection} channel is rather similar to the \textit{first CE ejection} one, though in this case companions are in the WD stage. The characteristics of this configuration imply that its final product should be systems with periods shorter than the ones found in previous configurations, and also that the range covered by the period distribution could be wider. Due to these properties, we compare this channel to our sdB candidates born after two CE episodes, the bottom panels in Fig. \ref{fig:set}. As can be understood from the distributions, we do find that most of the companions are COWDs, with a small contribution of ONeWDs and an even smaller fraction of NSs. A striking difference to the H03 predictions is that, although all our periods are shorter than the ones found in other channels and clustered around 0.05 days, they do not cover the expected wider range when compared to the first CE channel. In this scenario, it is possible to realize that some members of our 1CE episode channel actually match the definition of H03's \textit{second CE ejection}, in which case the period range being covered is indeed wider, going up as far as $\sim 100$ days. If this is the case, however, the period peak is slightly displaced towards longer periods at 0.1 days. Further, the peaks in mass distribution for the subset of stars that have experienced 2 CE episodes and have progenitor masses above (below) the MHeF limit match the predictions of H03, with a value of $\sim$0.33 ($\sim$ 0.46) M$_\odot$. The more massive progenitors (above MHeF) account for most of the systems found, in line with the description of H03.

H03 also include the \textit{merger} channel, which usually represents a non-negligible fraction of the total amount of the sdB candidates in all their populations (see their tables 1 and 2). Our mergers have not been analyzed through detailed models as what has been done in H02, which means that our results are referential and would only set an upper limit to the total number of sdBs being formed through this process (i.e., some of our mergers could be stars that do not ignite helium or follow a different evolutionary path). Nevertheless, we can see agreement in the mass range of our merger candidates.

We also note that figure 5 in H02 is similar to our Fig. \ref{fig:kiel_theo}, though the observational sample currently available is much richer and the hydrogen-rich layer treatment is different. Probably because of this reason, and the fact that we have not conducted a current day population study, the coverage within the Kiel diagram more readily matches the observations in our work. 

We briefly address a different study by \citet{Clausen2012}, who mention the relevance of hydrogen-rich shell in sdBs, but do not implement them within their rapid BPS code. Instead, they check whether their sdB candidates would remain in the $\log{g}$-$\log{T}$ box (chosen as their definition of sdBs) if those same stars had hydrogen-rich shells, based on results of \citet{Caloi1972}. However, their lower limit for surface gravity is set at $\log{g} = 5$, leaving a non-negligible fraction of the observational sample out of their parameter study (see Fig. \ref{fig:kiel_theo}).

Another important element raised by \citet{Clausen2012} is the mass cutoff for helium ignition. They consider both a 0 and 5\% cutoff value through their $f_{He}$ parameter, similar to what we have explored in our parameter variations. Accordingly, the impact in their population is evident, as can be concluded from comparing any pair of runs where only the $f_{He}$ is different in their table 1. For example their run 6 has about 15 times more resampled systems than run 7, which is potentially related to the spike that appears when considering a non-zero threshold in Fig. \ref{fig:metallicities}.

We note, however, that a direct one-to-one comparison is not possible at this point, particularly because our study does not predict a current-day population and therefore many of the conclusions drawn within \citet{Clausen2012} cannot be tested. We intend to present our estimates of the current-day sdB population in a following study.

\section{Conclusions}\label{sec:conclusion}

We have explored parameters influencing the different sdB formation channels and their characteristics as predicted by \textsc{compas}. Our results are in agreement with previous studies that have used population synthesis to analyze the properties of these stars, showing that \textsc{compas} and the prescription for hydrogen-shells are elements that further expand the set of tools available when analyzing sdBs. A more in-depth comparison with previous studies and the current day Galactic population of sdBs requires additional work, including a sampling based on the Galactic star formation rate and the different mass fractions and metallicities of each component of the Galaxy (Bulge, Disks and Halo), elements which we intend to analyze in a future publication. We foresee that this upcoming study will allow a comparison between our 1CE channel and e.g., \citet{Schaffenroth2022}, potentially providing an additional constraint to the parameters explored in the current work. Similarly, it will be possible to compare the total number of sdBs in the Galaxy at the current epoch as predicted by the \textsc{compas} BPS approach against the observational results of the 500pc sample \citep{Dawson2024}.

Moreover, the improved coverage of the \textit{sdB box} within the Kiel diagram highlights the potential of our prescription for sdBs with hydrogen-rich shells to further improve the predictions of rapid BPS models, particularly when it comes to comparing to observables linked to stellar radii and luminosity, such as the surface gravity and effective temperature distributions. An extended coverage of the models towards more massive sdBs would be ideal to further improve our predictions and avoid unnecessary mass cuts in the population, while a detailed study on the hydrogen-rich envelope relation with the different formation channels could impose additional constraints on the population. If each sub-channel presents a different hydrogen-envelope mass distribution, unlike the randomly sampled envelopes of our work, then we could check whether the corresponding number and characteristics of the sdBs born through each channel are in agreement with the observed Galactic sdB population or not. These potential differences could stem from the characteristics (stability) of the mass transfer episode that gives birth to a given sdB population.

We also note that it is still complicated to disentangle the different factors contributing to the observed characteristics of each analyzed physical property, in cases such as the period and mass distributions of the different formation channels. Moreover, additional parameters not explored in this work could reveal yet more properties and/or constraints for the evolution of sdBs, such as the case of magnetic braking \citep[e.g.,][]{Blomberg2024} and companions below the hydrogen burning limit \citep[Brown Dwarfs, see e.g.,][]{Schaffenroth2022}. We highlight, however, that the companion mass distribution could potentially constrain the angular momentum loss and critical mass prescriptions. Similarly, the mass distribution of sdBs with periods in the range $\sim 1$ -- 10 days could help to further test the threshold at which naked cores with masses below the expected helium-core mass at helium ignition would indeed ignite and become sdBs.

Finally, our chosen model (not to be confused with a best model, since we have not compared with the current-day Galactic sdB population) predicts the existence of sdB plus NS systems being born through a very specific channel, which we labeled as ECE+RLOF. While the expected occurrence of these systems seems to be low, it is interesting that a big fraction of them are born from systems that experience an early CE (ECE) episode, but the sdB candidate is born from a late stable mass transfer event. Even though we also predict sdB + NS systems from the 2CE channel, the ECE+RLOF one should be $\sim 10$ times more common, with periods longer than 1 day. Another interesting type of systems that can be found in our results are COWD+sdB pairs from the 1CE or 2CE channel, which can be understood as potential gravitational wave sources (owing to their close orbits) and will be of interest for LISA. The formation rate of these events is also something that we expect to predict in a future publication where a Galaxy-like population will be studied.

\section*{Acknowledgements}

We thank the anonymous referee for their helpful comments. NRS thanks Evan Bauer for insightful discussions, and also acknowledges valuable support received from the \textsc{compas} team, particularly from Reinhold Willcox, Ilya Mandel, Jeff Riley and Alejandro Vigna-G\'omez. AJR acknowledges financial support from the Australian Research Council under award number FT170100243. 

\section*{Data Availability}

All data will be available upon reasonable request to the corresponding author.

\bibliographystyle{paslike}
\bibliography{main}

\begin{thebibliography}{}
\expandafter\ifx\csname natexlab\endcsname\relax\def\natexlab#1{#1}\fi
\providecommand{\url}[1]{\href{#1}{#1}}
\providecommand{\dodoi}[1]{doi:~\href{http://doi.org/#1}{\nolinkurl{#1}}}
\providecommand{\doeprint}[1]{\href{http://ascl.net/#1}{\nolinkurl{http://ascl.net/#1}}}
\providecommand{\doarXiv}[1]{\href{https://arxiv.org/abs/#1}{\nolinkurl{https://arxiv.org/abs/#1}}}

\bibitem[{{Amaro-Seoane} {et~al.}(2023){Amaro-Seoane}, {Andrews}, {Arca Sedda}, {Askar}, {Baghi}, {Balasov}, {Bartos}, {Bavera}, {Bellovary}, {Berry}, {Berti}, {Bianchi}, {Blecha}, {Blondin}, {Bogdanovi{\'c}}, {Boissier}, {Bonetti}, {Bonoli}, {Bortolas}, {Breivik}, {Capelo}, {Caramete}, {Cattorini}, {Charisi}, {Chaty}, {Chen}, {Chru{\'s}li{\'n}ska}, {Chua}, {Church}, {Colpi}, {D'Orazio}, {Danielski}, {Davies}, {Dayal}, {De Rosa}, {Derdzinski}, {Destounis}, {Dotti}, {Dutan}, {Dvorkin}, {Fabj}, {Foglizzo}, {Ford}, {Fouvry}, {Franchini}, {Fragos}, {Fryer}, {Gaspari}, {Gerosa}, {Graziani}, {Groot}, {Habouzit}, {Haggard}, {Haiman}, {Han}, {Istrate}, {Johansson}, {Khan}, {Kimpson}, {Kokkotas}, {Kong}, {Korol}, {Kremer}, {Kupfer}, {Lamberts}, {Larson}, {Lau}, {Liu}, {Lloyd-Ronning}, {Lodato}, {Lupi}, {Ma}, {Maccarone}, {Mandel}, {Mangiagli}, {Mapelli}, {Mathis}, {Mayer}, {McGee}, {McKernan}, {Miller}, {Mota}, {Mumpower}, {Nasim}, {Nelemans}, {Noble}, {Pacucci}, {Panessa}, {Paschalidis}, {Pfister}, {Porquet},
  {Quenby}, {Ricarte}, {R{\"o}pke}, {Regan}, {Rosswog}, {Ruiter}, {Ruiz}, {Runnoe}, {Schneider}, {Schnittman}, {Secunda}, {Sesana}, {Seto}, {Shao}, {Shapiro}, {Sopuerta}, {Stone}, {Suvorov}, {Tamanini}, {Tamfal}, {Tauris}, {Temmink}, {Tomsick}, {Toonen}, {Torres-Orjuela}, {Toscani}, {Tsokaros}, {Unal}, {V{\'a}zquez-Aceves}, {Valiante}, {van Putten}, {van Roestel}, {Vignali}, {Volonteri}, {Wu}, {Younsi}, {Yu}, {Zane}, {Zwick}, {Antonini}, {Baibhav}, {Barausse}, {Bonilla Rivera}, {Branchesi}, {Branduardi-Raymont}, {Burdge}, {Chakraborty}, {Cuadra}, {Dage}, {Davis}, {de Mink}, {Decarli}, {Doneva}, {Escoffier}, {Gandhi}, {Haardt}, {Lousto}, {Nissanke}, {Nordhaus}, {O'Shaughnessy}, {Portegies Zwart}, {Pound}, {Schussler}, {Sergijenko}, {Spallicci}, {Vernieri}, \& {Vigna-G{\'o}mez}}]{AmaroSeoane2023}
{Amaro-Seoane}, P., {Andrews}, J., {Arca Sedda}, M., {et~al.} 2023, Living Reviews in Relativity, 26, 2, \dodoi{10.1007/s41114-022-00041-y}

\bibitem[{{Arancibia-Rojas} {et~al.}(2024){Arancibia-Rojas}, {Zorotovic}, {Vu{\v{c}}kovi{\'c}}, {Bobrick}, {Vos}, \& {Piraino-Cerda}}]{ArancibiaRojas2024}
{Arancibia-Rojas}, E., {Zorotovic}, M., {Vu{\v{c}}kovi{\'c}}, M., {et~al.} 2024, \mnras, 527, 11184, \dodoi{10.1093/mnras/stad3891}

\bibitem[{{Asplund} {et~al.}(2009){Asplund}, {Grevesse}, {Sauval}, \& {Scott}}]{Asplund2009}
{Asplund}, M., {Grevesse}, N., {Sauval}, A.~J., \& {Scott}, P. 2009, \araa, 47, 481, \dodoi{10.1146/annurev.astro.46.060407.145222}

\bibitem[{{Bauer} \& {Kupfer}(2021)}]{Bauer2021}
{Bauer}, E.~B., \& {Kupfer}, T. 2021, \apj, 922, 245, \dodoi{10.3847/1538-4357/ac25f0}

\bibitem[{{Belczynski} {et~al.}(2008){Belczynski}, {Kalogera}, {Rasio}, {Taam}, {Zezas}, {Bulik}, {Maccarone}, \& {Ivanova}}]{Belczynski2008}
{Belczynski}, K., {Kalogera}, V., {Rasio}, F.~A., {et~al.} 2008, \apjs, 174, 223, \dodoi{10.1086/521026}

\bibitem[{{Bertelli} {et~al.}(1994){Bertelli}, {Bressan}, {Chiosi}, {Fagotto}, \& {Nasi}}]{Bertelli1994}
{Bertelli}, G., {Bressan}, A., {Chiosi}, C., {Fagotto}, F., \& {Nasi}, E. 1994, \aaps, 106, 275

\bibitem[{{Blomberg} {et~al.}(2024){Blomberg}, {El-Badry}, \& {Breivik}}]{Blomberg2024}
{Blomberg}, L., {El-Badry}, K., \& {Breivik}, K. 2024, arXiv e-prints, arXiv:2408.15334, \dodoi{10.48550/arXiv.2408.15334}

\bibitem[{{Brassard} {et~al.}(2001){Brassard}, {Fontaine}, {Bill{\`e}res}, {Charpinet}, {Liebert}, \& {Saffer}}]{Brassard2001}
{Brassard}, P., {Fontaine}, G., {Bill{\`e}res}, M., {et~al.} 2001, \apj, 563, 1013, \dodoi{10.1086/323959}

\bibitem[{{Breivik} {et~al.}(2020){Breivik}, {Coughlin}, {Zevin}, {Rodriguez}, {Kremer}, {Ye}, {Andrews}, {Kurkowski}, {Digman}, {Larson}, \& {Rasio}}]{Breivik2020}
{Breivik}, K., {Coughlin}, S., {Zevin}, M., {et~al.} 2020, \apj, 898, 71, \dodoi{10.3847/1538-4357/ab9d85}

\bibitem[{{Broekgaarden} {et~al.}(2022){Broekgaarden}, {Berger}, {Stevenson}, {Justham}, {Mandel}, {Chru{\'s}li{\'n}ska}, {van Son}, {Wagg}, {Vigna-G{\'o}mez}, {de Mink}, {Chattopadhyay}, \& {Neijssel}}]{Broekgaarden2022}
{Broekgaarden}, F.~S., {Berger}, E., {Stevenson}, S., {et~al.} 2022, \mnras, 516, 5737, \dodoi{10.1093/mnras/stac1677}

\bibitem[{{Brown}(2004)}]{Brown2004}
{Brown}, T.~M. 2004, \apss, 291, 215, \dodoi{10.1023/B:ASTR.0000044324.60467.d2}

\bibitem[{{Caloi}(1972)}]{Caloi1972}
{Caloi}, V. 1972, \aap, 20, 357

\bibitem[{{Cassisi}(2014)}]{Cassisi2014}
{Cassisi}, S. 2014, in EAS Publications Series, Vol.~65, EAS Publications Series, ed. Y.~{Lebreton}, D.~{Valls-Gabaud}, \& C.~{Charbonnel}, 17--74, \dodoi{10.1051/eas/1465002}

\bibitem[{{Cassisi} {et~al.}(2016){Cassisi}, {Salaris}, \& {Pietrinferni}}]{Cassisi2016}
{Cassisi}, S., {Salaris}, M., \& {Pietrinferni}, A. 2016, \aap, 585, A124, \dodoi{10.1051/0004-6361/201527412}

\bibitem[{{Chen} {et~al.}(2013){Chen}, {Han}, {Deca}, \& {Podsiadlowski}}]{Chen2013}
{Chen}, X., {Han}, Z., {Deca}, J., \& {Podsiadlowski}, P. 2013, \mnras, 434, 186, \dodoi{10.1093/mnras/stt992}

\bibitem[{{Choi} {et~al.}(2016){Choi}, {Dotter}, {Conroy}, {Cantiello}, {Paxton}, \& {Johnson}}]{Choi2016}
{Choi}, J., {Dotter}, A., {Conroy}, C., {et~al.} 2016, \apj, 823, 102, \dodoi{10.3847/0004-637X/823/2/102}

\bibitem[{{Clausen} {et~al.}(2012){Clausen}, {Wade}, {Kopparapu}, \& {O'Shaughnessy}}]{Clausen2012}
{Clausen}, D., {Wade}, R.~A., {Kopparapu}, R.~K., \& {O'Shaughnessy}, R. 2012, \apj, 746, 186, \dodoi{10.1088/0004-637X/746/2/186}

\bibitem[{{Culpan} {et~al.}(2022){Culpan}, {Geier}, {Reindl}, {Pelisoli}, {Gentile Fusillo}, \& {Vorontseva}}]{Culpan2022}
{Culpan}, R., {Geier}, S., {Reindl}, N., {et~al.} 2022, \aap, 662, A40, \dodoi{10.1051/0004-6361/202243337}

\bibitem[{{Dawson} {et~al.}(2024){Dawson}, {Geier}, {Heber}, {Pelisoli}, {Dorsch}, {Schaffenroth}, {Reindl}, {Culpan}, {Pritzkuleit}, {Vos}, {Soemitro}, {Roth}, {Schneider}, {Uzundag}, {Vu{\v{c}}kovi{\'c}}, {Antunes Amaral}, {Istrate}, {Justham}, {{\O}stensen}, {Telting}, {Djupvik}, {Raddi}, {Green}, {Jeffery}, {Kepler}, {Munday}, {Steinmetz}, \& {Kupfer}}]{Dawson2024}
{Dawson}, H., {Geier}, S., {Heber}, U., {et~al.} 2024, \aap, 686, A25, \dodoi{10.1051/0004-6361/202348319}

\bibitem[{{D'Cruz} {et~al.}(1996){D'Cruz}, {Dorman}, {Rood}, \& {O'Connell}}]{DCruz1996}
{D'Cruz}, N.~L., {Dorman}, B., {Rood}, R.~T., \& {O'Connell}, R.~W. 1996, \apj, 466, 359, \dodoi{10.1086/177515}

\bibitem[{{de Kool}(1990)}]{deKool1990}
{de Kool}, M. 1990, \apj, 358, 189, \dodoi{10.1086/168974}

\bibitem[{{De Marco} {et~al.}(2011){De Marco}, {Passy}, {Moe}, {Herwig}, {Mac Low}, \& {Paxton}}]{DeMarco2011}
{De Marco}, O., {Passy}, J.-C., {Moe}, M., {et~al.} 2011, \mnras, 411, 2277, \dodoi{10.1111/j.1365-2966.2010.17891.x}

\bibitem[{{Ge} {et~al.}(2020){Ge}, {Webbink}, {Chen}, \& {Han}}]{Ge2020}
{Ge}, H., {Webbink}, R.~F., {Chen}, X., \& {Han}, Z. 2020, \apj, 899, 132, \dodoi{10.3847/1538-4357/aba7b7}

\bibitem[{{Geier} {et~al.}(2013){Geier}, {Marsh}, {Wang}, {Dunlap}, {Barlow}, {Schaffenroth}, {Chen}, {Irrgang}, {Maxted}, {Ziegerer}, {Kupfer}, {Miszalski}, {Heber}, {Han}, {Shporer}, {Telting}, {G{\"a}nsicke}, {{\O}stensen}, {O'Toole}, \& {Napiwotzki}}]{Geier2013}
{Geier}, S., {Marsh}, T.~R., {Wang}, B., {et~al.} 2013, \aap, 554, A54, \dodoi{10.1051/0004-6361/201321395}

\bibitem[{{Ghasemi} {et~al.}(2017){Ghasemi}, {Moravveji}, {Aerts}, {Safari}, \& {Vu{\v{c}}kovi{\'c}}}]{Ghasemi2017}
{Ghasemi}, H., {Moravveji}, E., {Aerts}, C., {Safari}, H., \& {Vu{\v{c}}kovi{\'c}}, M. 2017, \mnras, 465, 1518, \dodoi{10.1093/mnras/stw2839}

\bibitem[{{Hall} \& {Jeffery}(2016)}]{Hall2016}
{Hall}, P.~D., \& {Jeffery}, C.~S. 2016, \mnras, 463, 2756, \dodoi{10.1093/mnras/stw2188}

\bibitem[{{Han} {et~al.}(2003){Han}, {Podsiadlowski}, {Maxted}, \& {Marsh}}]{Han2003}
{Han}, Z., {Podsiadlowski}, P., {Maxted}, P.~F.~L., \& {Marsh}, T.~R. 2003, \mnras, 341, 669, \dodoi{10.1046/j.1365-8711.2003.06451.x}

\bibitem[{{Han} {et~al.}(2002){Han}, {Podsiadlowski}, {Maxted}, {Marsh}, \& {Ivanova}}]{Han2002}
{Han}, Z., {Podsiadlowski}, P., {Maxted}, P.~F.~L., {Marsh}, T.~R., \& {Ivanova}, N. 2002, \mnras, 336, 449, \dodoi{10.1046/j.1365-8711.2002.05752.x}

\bibitem[{{Heber}(2016)}]{Heber2016}
{Heber}, U. 2016, \pasp, 128, 082001, \dodoi{10.1088/1538-3873/128/966/082001}

\bibitem[{{Hjellming} \& {Webbink}(1987)}]{Hjellming1987}
{Hjellming}, M.~S., \& {Webbink}, R.~F. 1987, \apj, 318, 794, \dodoi{10.1086/165412}

\bibitem[{{Hurley} {et~al.}(2000){Hurley}, {Pols}, \& {Tout}}]{Hurley2000}
{Hurley}, J.~R., {Pols}, O.~R., \& {Tout}, C.~A. 2000, \mnras, 315, 543, \dodoi{10.1046/j.1365-8711.2000.03426.x}

\bibitem[{{Hurley} {et~al.}(2002){Hurley}, {Tout}, \& {Pols}}]{Hurley2002}
{Hurley}, J.~R., {Tout}, C.~A., \& {Pols}, O.~R. 2002, \mnras, 329, 897, \dodoi{10.1046/j.1365-8711.2002.05038.x}

\bibitem[{{Iben} \& {Tutukov}(1987)}]{Iben1987}
{Iben}, Icko, J., \& {Tutukov}, A.~V. 1987, \apj, 313, 727, \dodoi{10.1086/165011}

\bibitem[{{Ivanova} {et~al.}(2020){Ivanova}, {Justham}, \& {Ricker}}]{Ivanova2020}
{Ivanova}, N., {Justham}, S., \& {Ricker}, P. 2020, {Common Envelope Evolution}, \dodoi{10.1088/2514-3433/abb6f0}

\bibitem[{{Izzard} {et~al.}(2004){Izzard}, {Tout}, {Karakas}, \& {Pols}}]{Izzard2004}
{Izzard}, R.~G., {Tout}, C.~A., {Karakas}, A.~I., \& {Pols}, O.~R. 2004, \mnras, 350, 407, \dodoi{10.1111/j.1365-2966.2004.07446.x}

\bibitem[{{Krzesinski} {et~al.}(2014){Krzesinski}, {Blokesz}, {Baran}, \& {Bachulski}}]{Krzesinski2014}
{Krzesinski}, J., {Blokesz}, A., {Baran}, A.~S., \& {Bachulski}, S. 2014, \actaa, 64, 151

\bibitem[{{Kupfer} {et~al.}(2018){Kupfer}, {Korol}, {Shah}, {Nelemans}, {Marsh}, {Ramsay}, {Groot}, {Steeghs}, \& {Rossi}}]{Kupfer2018}
{Kupfer}, T., {Korol}, V., {Shah}, S., {et~al.} 2018, \mnras, 480, 302, \dodoi{10.1093/mnras/sty1545}

\bibitem[{{Kupfer} {et~al.}(2022){Kupfer}, {Bauer}, {van Roestel}, {Bellm}, {Bildsten}, {Fuller}, {Prince}, {Heber}, {Geier}, {Green}, {Kulkarni}, {Bloemen}, {Laher}, {Rusholme}, \& {Schneider}}]{Kupfer2022}
{Kupfer}, T., {Bauer}, E.~B., {van Roestel}, J., {et~al.} 2022, \apjl, 925, L12, \dodoi{10.3847/2041-8213/ac48f1}

\bibitem[{{Lei} {et~al.}(2023){Lei}, {He}, {N{\'e}meth}, {Vos}, {Zou}, {Hu}, {Xiao}, {Yan}, \& {Zhao}}]{Lei2023a}
{Lei}, Z., {He}, R., {N{\'e}meth}, P., {et~al.} 2023, \apj, 942, 109, \dodoi{10.3847/1538-4357/aca542}

\bibitem[{{Livne}(1990)}]{Livne1990}
{Livne}, E. 1990, \apjl, 354, L53, \dodoi{10.1086/185721}

\bibitem[{{Maxted} {et~al.}(2001){Maxted}, {Heber}, {Marsh}, \& {North}}]{Maxted2001}
{Maxted}, P.~F.~L., {Heber}, U., {Marsh}, T.~R., \& {North}, R.~C. 2001, \mnras, 326, 1391, \dodoi{10.1111/j.1365-2966.2001.04714.x}

\bibitem[{{Moe} \& {Di Stefano}(2017)}]{Moe2017}
{Moe}, M., \& {Di Stefano}, R. 2017, \apjs, 230, 15, \dodoi{10.3847/1538-4365/aa6fb6}

\bibitem[{{Nelemans}(2010)}]{Nelemans2010}
{Nelemans}, G. 2010, \apss, 329, 25, \dodoi{10.1007/s10509-010-0392-0}

\bibitem[{{Neunteufel} {et~al.}(2016){Neunteufel}, {Yoon}, \& {Langer}}]{Neunteufel2016}
{Neunteufel}, P., {Yoon}, S.~C., \& {Langer}, N. 2016, \aap, 589, A43, \dodoi{10.1051/0004-6361/201527845}

\bibitem[{{Nomoto}(1982)}]{Nomoto1982a}
{Nomoto}, K. 1982, \apj, 253, 798, \dodoi{10.1086/159682}

\bibitem[{{Ostrowski} {et~al.}(2021){Ostrowski}, {Baran}, {Sanjayan}, \& {Sahoo}}]{Ostrowski2021}
{Ostrowski}, J., {Baran}, A.~S., {Sanjayan}, S., \& {Sahoo}, S.~K. 2021, \mnras, 503, 4646, \dodoi{10.1093/mnras/staa3751}

\bibitem[{{Paczy{\'n}ski} \& {Sienkiewicz}(1972)}]{Paczynski1972}
{Paczy{\'n}ski}, B., \& {Sienkiewicz}, R. 1972, \actaa, 22, 73

\bibitem[{{Paxton} {et~al.}(2011){Paxton}, {Bildsten}, {Dotter}, {Herwig}, {Lesaffre}, \& {Timmes}}]{Paxton2011}
{Paxton}, B., {Bildsten}, L., {Dotter}, A., {et~al.} 2011, \apjs, 192, 3, \dodoi{10.1088/0067-0049/192/1/3}

\bibitem[{{Paxton} {et~al.}(2013){Paxton}, {Cantiello}, {Arras}, {Bildsten}, {Brown}, {Dotter}, {Mankovich}, {Montgomery}, {Stello}, {Timmes}, \& {Townsend}}]{Paxton2013}
{Paxton}, B., {Cantiello}, M., {Arras}, P., {et~al.} 2013, \apjs, 208, 4, \dodoi{10.1088/0067-0049/208/1/4}

\bibitem[{{Paxton} {et~al.}(2015){Paxton}, {Marchant}, {Schwab}, {Bauer}, {Bildsten}, {Cantiello}, {Dessart}, {Farmer}, {Hu}, {Langer}, {Townsend}, {Townsley}, \& {Timmes}}]{Paxton2015}
{Paxton}, B., {Marchant}, P., {Schwab}, J., {et~al.} 2015, \apjs, 220, 15, \dodoi{10.1088/0067-0049/220/1/15}

\bibitem[{{Paxton} {et~al.}(2018){Paxton}, {Schwab}, {Bauer}, {Bildsten}, {Blinnikov}, {Duffell}, {Farmer}, {Goldberg}, {Marchant}, {Sorokina}, {Thoul}, {Townsend}, \& {Timmes}}]{Paxton2018}
{Paxton}, B., {Schwab}, J., {Bauer}, E.~B., {et~al.} 2018, \apjs, 234, 34, \dodoi{10.3847/1538-4365/aaa5a8}

\bibitem[{{Paxton} {et~al.}(2019){Paxton}, {Smolec}, {Schwab}, {Gautschy}, {Bildsten}, {Cantiello}, {Dotter}, {Farmer}, {Goldberg}, {Jermyn}, {Kanbur}, {Marchant}, {Thoul}, {Townsend}, {Wolf}, {Zhang}, \& {Timmes}}]{Paxton2019}
{Paxton}, B., {Smolec}, R., {Schwab}, J., {et~al.} 2019, \apjs, 243, 10, \dodoi{10.3847/1538-4365/ab2241}

\bibitem[{{Pelisoli} {et~al.}(2020){Pelisoli}, {Vos}, {Geier}, {Schaffenroth}, \& {Baran}}]{Pelisoli2020}
{Pelisoli}, I., {Vos}, J., {Geier}, S., {Schaffenroth}, V., \& {Baran}, A.~S. 2020, \aap, 642, A180, \dodoi{10.1051/0004-6361/202038473}

\bibitem[{{Peters}(1964)}]{Peters1964}
{Peters}, P.~C. 1964, Physical Review, 136, 1224, \dodoi{10.1103/PhysRev.136.B1224}

\bibitem[{{Pietrukowicz} {et~al.}(2017){Pietrukowicz}, {Dziembowski}, {Latour}, {Angeloni}, {Poleski}, {di Mille}, {Soszy{\'n}ski}, {Udalski}, {Szyma{\'n}ski}, {Wyrzykowski}, {Koz{\l}owski}, {Skowron}, {Skowron}, {Mr{\'o}z}, {Pawlak}, \& {Ulaczyk}}]{Pietrukowicz2017}
{Pietrukowicz}, P., {Dziembowski}, W.~A., {Latour}, M., {et~al.} 2017, Nature Astronomy, 1, 0166, \dodoi{10.1038/s41550-017-0166}

\bibitem[{{Price} {et~al.}(2018){Price}, {Wurster}, {Tricco}, {Nixon}, {Toupin}, {Pettitt}, {Chan}, {Mentiplay}, {Laibe}, {Glover}, {Dobbs}, {Nealon}, {Liptai}, {Worpel}, {Bonnerot}, {Dipierro}, {Ballabio}, {Ragusa}, {Federrath}, {Iaconi}, {Reichardt}, {Forgan}, {Hutchison}, {Constantino}, {Ayliffe}, {Hirsh}, \& {Lodato}}]{Price2018}
{Price}, D.~J., {Wurster}, J., {Tricco}, T.~S., {et~al.} 2018, \pasa, 35, e031, \dodoi{10.1017/pasa.2018.25}

\bibitem[{{Reed} {et~al.}(2021){Reed}, {Slayton}, {Baran}, {Telting}, {{\O}stensen}, {Jeffery}, {Uzundag}, \& {Sanjayan}}]{Reed2021}
{Reed}, M.~D., {Slayton}, A., {Baran}, A.~S., {et~al.} 2021, \mnras, 507, 4178, \dodoi{10.1093/mnras/stab2405}

\bibitem[{{Renzini}(2023)}]{Renzini2023}
{Renzini}, A. 2023, \mnras, 521, 524, \dodoi{10.1093/mnras/stad159}

\bibitem[{{Riley} {et~al.}(2022){Riley}, {Agrawal}, {Barrett}, {Boyett}, {Broekgaarden}, {Chattopadhyay}, {Gaebel}, {Gittins}, {Hirai}, {Howitt}, {Justham}, {Khandelwal}, {Kummer}, {Lau}, {Mandel}, {de Mink}, {Neijssel}, {Riley}, {van Son}, {Stevenson}, {Vigna-G{\'o}mez}, {Vinciguerra}, {Wagg}, {Willcox}, \& {Team Compas}}]{Riley2022}
{Riley}, J., {Agrawal}, P., {Barrett}, J.~W., {et~al.} 2022, \apjs, 258, 34, \dodoi{10.3847/1538-4365/ac416c}

\bibitem[{{R{\"o}pke} \& {De Marco}(2023)}]{Roepke2023}
{R{\"o}pke}, F.~K., \& {De Marco}, O. 2023, Living Reviews in Computational Astrophysics, 9, 2, \dodoi{10.1007/s41115-023-00017-x}

\bibitem[{{Sargent} \& {Searle}(1968)}]{Sargent1968}
{Sargent}, W. L.~W., \& {Searle}, L. 1968, \apj, 152, 443, \dodoi{10.1086/149561}

\bibitem[{{Savonije} {et~al.}(1986){Savonije}, {de Kool}, \& {van den Heuvel}}]{Savonije1986}
{Savonije}, G.~J., {de Kool}, M., \& {van den Heuvel}, E.~P.~J. 1986, \aap, 155, 51

\bibitem[{{Schaffenroth} {et~al.}(2022){Schaffenroth}, {Pelisoli}, {Barlow}, {Geier}, \& {Kupfer}}]{Schaffenroth2022}
{Schaffenroth}, V., {Pelisoli}, I., {Barlow}, B.~N., {Geier}, S., \& {Kupfer}, T. 2022, \aap, 666, A182, \dodoi{10.1051/0004-6361/202244214}

\bibitem[{{Soberman} {et~al.}(1997){Soberman}, {Phinney}, \& {van den Heuvel}}]{Soberman1997}
{Soberman}, G.~E., {Phinney}, E.~S., \& {van den Heuvel}, E.~P.~J. 1997, \aap, 327, 620, \dodoi{10.48550/arXiv.astro-ph/9703016}

\bibitem[{{Stark} \& {Wade}(2003)}]{Stark2003}
{Stark}, M.~A., \& {Wade}, R.~A. 2003, \aj, 126, 1455, \dodoi{10.1086/377017}

\bibitem[{{Stevenson} \& {Clarke}(2022)}]{Stevenson2022}
{Stevenson}, S., \& {Clarke}, T.~A. 2022, \mnras, 517, 4034, \dodoi{10.1093/mnras/stac2936}

\bibitem[{{Sweigart} \& {Gross}(1978)}]{Sweigart1978}
{Sweigart}, A.~V., \& {Gross}, P.~G. 1978, \apjs, 36, 405, \dodoi{10.1086/190506}

\bibitem[{{Toonen} {et~al.}(2014){Toonen}, {Claeys}, {Mennekens}, \& {Ruiter}}]{Toonen2014}
{Toonen}, S., {Claeys}, J.~S.~W., {Mennekens}, N., \& {Ruiter}, A.~J. 2014, \aap, 562, A14, \dodoi{10.1051/0004-6361/201321576}

\bibitem[{{Toonen} {et~al.}(2012){Toonen}, {Nelemans}, \& {Portegies Zwart}}]{Toonen2012}
{Toonen}, S., {Nelemans}, G., \& {Portegies Zwart}, S. 2012, \aap, 546, A70, \dodoi{10.1051/0004-6361/201218966}

\bibitem[{{Tutukov} {et~al.}(1985){Tutukov}, {Fedorova}, {Ergma}, \& {Yungelson}}]{Tutukov1985}
{Tutukov}, A.~V., {Fedorova}, A.~V., {Ergma}, E.~V., \& {Yungelson}, L.~R. 1985, Soviet Astronomy Letters, 11, 52

\bibitem[{Virtanen {et~al.}(2020)Virtanen, Gommers, Oliphant, Haberland, Reddy, Cournapeau, Burovski, Peterson, Weckesser, Bright, {van der Walt}, Brett, Wilson, Millman, Mayorov, Nelson, Jones, Kern, Larson, Carey, Polat, Feng, Moore, {VanderPlas}, Laxalde, Perktold, Cimrman, Henriksen, Quintero, Harris, Archibald, Ribeiro, Pedregosa, {van Mulbregt}, \& {SciPy 1.0 Contributors}}]{Virtanen2020}
Virtanen, P., Gommers, R., Oliphant, T.~E., {et~al.} 2020, Nature Methods, 17, 261, \dodoi{10.1038/s41592-019-0686-2}

\bibitem[{{Vos} {et~al.}(2020){Vos}, {Bobrick}, \& {Vu{\v{c}}kovi{\'c}}}]{Vos2020}
{Vos}, J., {Bobrick}, A., \& {Vu{\v{c}}kovi{\'c}}, M. 2020, \aap, 641, A163, \dodoi{10.1051/0004-6361/201937195}

\bibitem[{{Wagg} {et~al.}(2022){Wagg}, {Broekgaarden}, {de Mink}, {Frankel}, {van Son}, \& {Justham}}]{Wagg2022}
{Wagg}, T., {Broekgaarden}, F.~S., {de Mink}, S.~E., {et~al.} 2022, \apj, 937, 118, \dodoi{10.3847/1538-4357/ac8675}

\bibitem[{{Webbink}(1984)}]{Webbink1984}
{Webbink}, R.~F. 1984, \apj, 277, 355, \dodoi{10.1086/161701}

\bibitem[{{Willcox} {et~al.}(2023){Willcox}, {MacLeod}, {Mandel}, \& {Hirai}}]{Willcox2023}
{Willcox}, R., {MacLeod}, M., {Mandel}, I., \& {Hirai}, R. 2023, \apj, 958, 138, \dodoi{10.3847/1538-4357/acffb1}

\bibitem[{{Woods} {et~al.}(2012){Woods}, {Ivanova}, {van der Sluys}, \& {Chaichenets}}]{Woods2012}
{Woods}, T.~E., {Ivanova}, N., {van der Sluys}, M.~V., \& {Chaichenets}, S. 2012, \apj, 744, 12, \dodoi{10.1088/0004-637X/744/1/12}

\bibitem[{{Woosley} \& {Weaver}(1994)}]{Woosley1994}
{Woosley}, S.~E., \& {Weaver}, T.~A. 1994, \apj, 423, 371, \dodoi{10.1086/173813}

\bibitem[{{Xiong} {et~al.}(2022){Xiong}, {Casagrande}, {Chen}, {Vos}, {Zhang}, {Justham}, {Li}, {Wu}, {Li}, \& {Han}}]{Xiong2022}
{Xiong}, H., {Casagrande}, L., {Chen}, X., {et~al.} 2022, \aap, 668, A112, \dodoi{10.1051/0004-6361/202244571}

\bibitem[{{Xu} \& {Li}(2010{\natexlab{a}})}]{Xu2010}
{Xu}, X.-J., \& {Li}, X.-D. 2010{\natexlab{a}}, \apj, 716, 114, \dodoi{10.1088/0004-637X/716/1/114}

\bibitem[{{Xu} \& {Li}(2010{\natexlab{b}})}]{Xu2010a}
---. 2010{\natexlab{b}}, \apj, 722, 1985, \dodoi{10.1088/0004-637X/722/2/1985}

\bibitem[{{Yungelson}(2008)}]{Yungelson2008}
{Yungelson}, L.~R. 2008, Astronomy Letters, 34, 620, \dodoi{10.1134/S1063773708090053}

\bibitem[{{Zorotovic} \& {Schreiber}(2013)}]{Zorotovic2013}
{Zorotovic}, M., \& {Schreiber}, M.~R. 2013, \aap, 549, A95, \dodoi{10.1051/0004-6361/201220321}

\bibitem[{{Zorotovic} {et~al.}(2010){Zorotovic}, {Schreiber}, {G{\"a}nsicke}, \& {Nebot G{\'o}mez-Mor{\'a}n}}]{Zorotovic2010}
{Zorotovic}, M., {Schreiber}, M.~R., {G{\"a}nsicke}, B.~T., \& {Nebot G{\'o}mez-Mor{\'a}n}, A. 2010, \aap, 520, A86, \dodoi{10.1051/0004-6361/200913658}

\end{thebibliography}

\begin{appendix}
\renewcommand\thefigure{\thesection.\arabic{figure}}
\renewcommand\thetable{\thesection.\arabic{table}}
\section{Complete List Of Configurations}\label{app:runs}

The specific configuration for each run is shown in Table \ref{tab:conf}. Even though there are some degeneracies due to no angular momentum loss at full accretion, we have kept all sets in order to compare similar numbers of candidates during each parameter variation.

\begin{table*}
\renewcommand{\arraystretch}{1.3}
    \caption{Parameters used on each \textsc{compas} configuration set. Proprieties not listed here were kept as default, or already specified in Table \ref{tab:params}.}\label{tab:conf}
    {\tablefont\begin{tabular}{@{\extracolsep{\fill}}lccccc|lccccc|lccccc}
        \toprule
        Set & $\alpha$ & q$_{\rm crit}$ & Z & FA & MLF & Set & $\alpha$ & q$_{\rm crit}$ & Z & FA & MLF & Set & $\alpha$ & q$_{\rm crit}$ & Z & FA & MLF \\ 
        \hline
        1 & 0.2 & NONE & 0.0012 & 0 & 0 & 55 & 1 & NONE & 0.0012 & 0 & 0 & 109 & 1.5 & NONE & 0.0012 & 0 & 0 \\
        2 & 0.2 & GE20 & 0.0012 & 0 & 0 & 56 & 1 & GE20 & 0.0012 & 0 & 0 & 110 & 1.5 & GE20 & 0.0012 & 0 & 0 \\
        3 & 0.2 & NONE & 0.0012 & 0.5 & 0 & 57 & 1 & NONE & 0.0012 & 0.5 & 0 & 111 & 1.5 & NONE & 0.0012 & 0.5 & 0 \\
        4 & 0.2 & GE20 & 0.0012 & 0.5 & 0 & 58 & 1 & GE20 & 0.0012 & 0.5 & 0 & 112 & 1.5 & GE20 & 0.0012 & 0.5 & 0 \\
        5 & 0.2 & NONE & 0.0012 & 1 & 0 & 59 & 1 & NONE & 0.0012 & 1 & 0 & 113 & 1.5 & NONE & 0.0012 & 1 & 0 \\
        6 & 0.2 & GE20 & 0.0012 & 1 & 0 & 60 & 1 & GE20 & 0.0012 & 1 & 0 & 114 & 1.5 & GE20 & 0.0012 & 1 & 0 \\
        7 & 0.2 & NONE & 0.0012 & 0 & 0.5 & 61 & 1 & NONE & 0.0012 & 0 & 0.5 & 115 & 1.5 & NONE & 0.0012 & 0 & 0.5 \\
        8 & 0.2 & GE20 & 0.0012 & 0 & 0.5 & 62 & 1 & GE20 & 0.0012 & 0 & 0.5 & 116 & 1.5 & GE20 & 0.0012 & 0 & 0.5 \\
        9 & 0.2 & NONE & 0.0012 & 0.5 & 0.5 & 63 & 1 & NONE & 0.0012 & 0.5 & 0.5 & 117 & 1.5 & NONE & 0.0012 & 0.5 & 0.5 \\
        10 & 0.2 & GE20 & 0.0012 & 0.5 & 0.5 & 64 & 1 & GE20 & 0.0012 & 0.5 & 0.5 & 118 & 1.5 & GE20 & 0.0012 & 0.5 & 0.5 \\
        11 & 0.2 & NONE & 0.0012 & 1 & 0.5 & 65 & 1 & NONE & 0.0012 & 1 & 0.5 & 119 & 1.5 & NONE & 0.0012 & 1 & 0.5 \\
        12 & 0.2 & GE20 & 0.0012 & 1 & 0.5 & 66 & 1 & GE20 & 0.0012 & 1 & 0.5 & 120 & 1.5 & GE20 & 0.0012 & 1 & 0.5 \\
        13 & 0.2 & NONE & 0.0012 & 0 & 1 & 67 & 1 & NONE & 0.0012 & 0 & 1 & 121 & 1.5 & NONE & 0.0012 & 0 & 1 \\
        14 & 0.2 & GE20 & 0.0012 & 0 & 1 & 68 & 1 & GE20 & 0.0012 & 0 & 1 & 122 & 1.5 & GE20 & 0.0012 & 0 & 1 \\
        15 & 0.2 & NONE & 0.0012 & 0.5 & 1 & 69 & 1 & NONE & 0.0012 & 0.5 & 1 & 123 & 1.5 & NONE & 0.0012 & 0.5 & 1 \\
        16 & 0.2 & GE20 & 0.0012 & 0.5 & 1 & 70 & 1 & GE20 & 0.0012 & 0.5 & 1 & 124 & 1.5 & GE20 & 0.0012 & 0.5 & 1 \\
        17 & 0.2 & NONE & 0.0012 & 1 & 1 & 71 & 1 & NONE & 0.0012 & 1 & 1 & 125 & 1.5 & NONE & 0.0012 & 1 & 1 \\
        18 & 0.2 & GE20 & 0.0012 & 1 & 1 & 72 & 1 & GE20 & 0.0012 & 1 & 1 & 126 & 1.5 & GE20 & 0.0012 & 1 & 1 \\
        19 & 0.2 & NONE & 0.0142 & 0 & 0 & 73 & 1 & NONE & 0.0142 & 0 & 0 & 127 & 1.5 & NONE & 0.0142 & 0 & 0 \\
        20 & 0.2 & GE20 & 0.0142 & 0 & 0 & 74 & 1 & GE20 & 0.0142 & 0 & 0 & 128 & 1.5 & GE20 & 0.0142 & 0 & 0 \\
        21 & 0.2 & NONE & 0.0142 & 0.5 & 0 & 75 & 1 & NONE & 0.0142 & 0.5 & 0 & 129 & 1.5 & NONE & 0.0142 & 0.5 & 0 \\
        22 & 0.2 & GE20 & 0.0142 & 0.5 & 0 & 76 & 1 & GE20 & 0.0142 & 0.5 & 0 & 130 & 1.5 & GE20 & 0.0142 & 0.5 & 0 \\
        23 & 0.2 & NONE & 0.0142 & 1 & 0 & 77 & 1 & NONE & 0.0142 & 1 & 0 & 131 & 1.5 & NONE & 0.0142 & 1 & 0 \\
        24 & 0.2 & GE20 & 0.0142 & 1 & 0 & 78 & 1 & GE20 & 0.0142 & 1 & 0 & 132 & 1.5 & GE20 & 0.0142 & 1 & 0 \\
        25 & 0.2 & NONE & 0.0142 & 0 & 0.5 & 79 & 1 & NONE & 0.0142 & 0 & 0.5 & 133 & 1.5 & NONE & 0.0142 & 0 & 0.5 \\
        26 & 0.2 & GE20 & 0.0142 & 0 & 0.5 & 80 & 1 & GE20 & 0.0142 & 0 & 0.5 & 134 & 1.5 & GE20 & 0.0142 & 0 & 0.5 \\
        27 & 0.2 & NONE & 0.0142 & 0.5 & 0.5 & 81 & 1 & NONE & 0.0142 & 0.5 & 0.5 & 135 & 1.5 & NONE & 0.0142 & 0.5 & 0.5 \\
        28 & 0.2 & GE20 & 0.0142 & 0.5 & 0.5 & 82 & 1 & GE20 & 0.0142 & 0.5 & 0.5 & 136 & 1.5 & GE20 & 0.0142 & 0.5 & 0.5 \\
        29 & 0.2 & NONE & 0.0142 & 1 & 0.5 & 83 & 1 & NONE & 0.0142 & 1 & 0.5 & 137 & 1.5 & NONE & 0.0142 & 1 & 0.5 \\
        30 & 0.2 & GE20 & 0.0142 & 1 & 0.5 & 84 & 1 & GE20 & 0.0142 & 1 & 0.5 & 138 & 1.5 & GE20 & 0.0142 & 1 & 0.5 \\
        31 & 0.2 & NONE & 0.0142 & 0 & 1 & 85 & 1 & NONE & 0.0142 & 0 & 1 & 139 & 1.5 & NONE & 0.0142 & 0 & 1 \\
        32 & 0.2 & GE20 & 0.0142 & 0 & 1 & 86 & 1 & GE20 & 0.0142 & 0 & 1 & 140 & 1.5 & GE20 & 0.0142 & 0 & 1 \\
        33 & 0.2 & NONE & 0.0142 & 0.5 & 1 & 87 & 1 & NONE & 0.0142 & 0.5 & 1 & 141 & 1.5 & NONE & 0.0142 & 0.5 & 1 \\
        34 & 0.2 & GE20 & 0.0142 & 0.5 & 1 & 88 & 1 & GE20 & 0.0142 & 0.5 & 1 & 142 & 1.5 & GE20 & 0.0142 & 0.5 & 1 \\
        35 & 0.2 & NONE & 0.0142 & 1 & 1 & 89 & 1 & NONE & 0.0142 & 1 & 1 & 143 & 1.5 & NONE & 0.0142 & 1 & 1 \\
        36 & 0.2 & GE20 & 0.0142 & 1 & 1 & 90 & 1 & GE20 & 0.0142 & 1 & 1 & 144 & 1.5 & GE20 & 0.0142 & 1 & 1 \\
        37 & 0.2 & NONE & 0.03 & 0 & 0 & 91 & 1 & NONE & 0.03 & 0 & 0 & 145 & 1.5 & NONE & 0.03 & 0 & 0 \\
        38 & 0.2 & GE20 & 0.03 & 0 & 0 & 92 & 1 & GE20 & 0.03 & 0 & 0 & 146 & 1.5 & GE20 & 0.03 & 0 & 0 \\
        39 & 0.2 & NONE & 0.03 & 0.5 & 0 & 93 & 1 & NONE & 0.03 & 0.5 & 0 & 147 & 1.5 & NONE & 0.03 & 0.5 & 0 \\
        40 & 0.2 & GE20 & 0.03 & 0.5 & 0 & 94 & 1 & GE20 & 0.03 & 0.5 & 0 & 148 & 1.5 & GE20 & 0.03 & 0.5 & 0 \\
        41 & 0.2 & NONE & 0.03 & 1 & 0 & 95 & 1 & NONE & 0.03 & 1 & 0 & 149 & 1.5 & NONE & 0.03 & 1 & 0 \\
        42 & 0.2 & GE20 & 0.03 & 1 & 0 & 96 & 1 & GE20 & 0.03 & 1 & 0 & 150 & 1.5 & GE20 & 0.03 & 1 & 0 \\
        43 & 0.2 & NONE & 0.03 & 0 & 0.5 & 97 & 1 & NONE & 0.03 & 0 & 0.5 & 151 & 1.5 & NONE & 0.03 & 0 & 0.5 \\
        44 & 0.2 & GE20 & 0.03 & 0 & 0.5 & 98 & 1 & GE20 & 0.03 & 0 & 0.5 & 152 & 1.5 & GE20 & 0.03 & 0 & 0.5 \\
        45 & 0.2 & NONE & 0.03 & 0.5 & 0.5 & 99 & 1 & NONE & 0.03 & 0.5 & 0.5 & 153 & 1.5 & NONE & 0.03 & 0.5 & 0.5 \\
        46 & 0.2 & GE20 & 0.03 & 0.5 & 0.5 & 100 & 1 & GE20 & 0.03 & 0.5 & 0.5 & 154 & 1.5 & GE20 & 0.03 & 0.5 & 0.5 \\
        47 & 0.2 & NONE & 0.03 & 1 & 0.5 & 101 & 1 & NONE & 0.03 & 1 & 0.5 & 155 & 1.5 & NONE & 0.03 & 1 & 0.5 \\
        48 & 0.2 & GE20 & 0.03 & 1 & 0.5 & 102 & 1 & GE20 & 0.03 & 1 & 0.5 & 156 & 1.5 & GE20 & 0.03 & 1 & 0.5 \\
        49 & 0.2 & NONE & 0.03 & 0 & 1 & 103 & 1 & NONE & 0.03 & 0 & 1 & 157 & 1.5 & NONE & 0.03 & 0 & 1 \\
        50 & 0.2 & GE20 & 0.03 & 0 & 1 & 104 & 1 & GE20 & 0.03 & 0 & 1 & 158 & 1.5 & GE20 & 0.03 & 0 & 1 \\
        51 & 0.2 & NONE & 0.03 & 0.5 & 1 & 105 & 1 & NONE & 0.03 & 0.5 & 1 & 159 & 1.5 & NONE & 0.03 & 0.5 & 1 \\
        52 & 0.2 & GE20 & 0.03 & 0.5 & 1 & 106 & 1 & GE20 & 0.03 & 0.5 & 1 & 160 & 1.5 & GE20 & 0.03 & 0.5 & 1 \\
        53 & 0.2 & NONE & 0.03 & 1 & 1 & 107 & 1 & NONE & 0.03 & 1 & 1 & 161 & 1.5 & NONE & 0.03 & 1 & 1 \\
        54 & 0.2 & GE20 & 0.03 & 1 & 1 & 108 & 1 & GE20 & 0.03 & 1 & 1 & 162 & 1.5 & GE20 & 0.03 & 1 & 1
    \botrule
	\end{tabular}}
\end{table*}

\section{Coefficients for Fits}\label{app:coefs}

\begin{table*}
	\caption{Coefficients for age fit, as presented in Equation \ref{eq:lifetime}.}\label{tab:A}
	{\tablefont\begin{tabular}{@{\extracolsep{\fill}}lcccccc}
		\toprule
		Progenitor Mass & $A_1$ & $A_2$ & $A_3$ & $A_4$ & $A_5$ \\
		\hline
        Below MHeF & 1.22188566 & 0.26004337 & -1.59709934 & 0.07264813 & -61.88868775 \\
        Above MHeF & 0.05161968 & 0.25380777 & 0.09981282  & 0.00252778 & 424.26395881
         \botrule
	\end{tabular}}
\end{table*}

\subsection{Lifetime}
The lifetime (maximum age) of a HeMS star with a hydrogen-rich shell can be computed by using Equation \ref{eq:lifetime}, and the different $A_i$ coefficients are shown in Table \ref{tab:A}.

\subsection{Radius}\label{ap:rad}
The coefficients of the fit shown in Equation \ref{eq:radius} can be computed using one of the following forms:

\begin{align}
    F_1 = \left(a + bM + cM^2\right)\left(d + eM_{\rm H} + fM^2\right) + g + hM_{\rm H} ,\\
    F_2 = \left(a + \frac{b}{0.323 - M + iM^2} + cM^2\right)\left(d + eM_{\rm H} + fM_{\rm H}^2\right) + g + hM ,\\
    F_3 =  F_1\left(a,b,c,d,e,f,g,h\right) + j\left(0.45 - M^3\right) + k\left(0.45 - M^4\right) + l\left(0.45 - M^5\right),
\end{align}

\noindent while the specific details of the $a,b,...,k,l$ constants for each $B_i$ coefficient are given in Table \ref{tab:B}.

\subsection{Luminosity}\label{ap:lum}
Similar to what was done for the stellar radius, the coefficients of the fit applied to luminosity and shown in Equation \ref{eq:luminosity} require one of the following:

\begin{align}
    F_4 = \left(a + bM + cM^2\right)\left(d+eM_{\rm H}\right) + f + gM_{\rm H},\\
    F_5 = \left(a +bM + \frac{c}{\left(M + d\right)^{e}}\right)\left(f + gM_{\rm H}\right) + h + iM_{\rm H},\\
    F_6 = \left(a + bM + cM^2 + dM^5\right)\left(e + fM_{\rm H}\right) + g + hM_{\rm H},
\end{align}

\noindent with specific values corresponding to the $a,b,...,h,i$ constants required for each $C_i$ being shown in Table \ref{tab:C}.

\begin{landscape}
\begin{table}
    \centering
	\caption{\centering Coefficients used in the radius fit, as presented in Equation \ref{eq:radius}. The specific form of each $F_i$ expression is defined in appendix \ref{ap:rad}.}\label{tab:B}
	{\tablefont\begin{tabular}{@{\extracolsep{\fill}}lc}
		\toprule
        \multicolumn{2}{c}{Progenitor mass below MHeF}\\
        \hline
		Coefficient & Expression \\
		\hline
        $B_1$ & $F_1\left(-0.434436, 1.973170, -1.205341, 0.313716, 370.599542, -77521.575181, 0.036542, 11.479660\right)$\\
        $B_2$ & $F_2\left(15.270717, 0.200441, 18.522382, 0.396108, 2.937080, -403.135079, -5.387696, -6.402536, 1.135359\right)$\\
        $B_3$ & $F_3\left(104.818378, -59.288148, 92.739946, 97.627387, 485.083943, 2762.327494, -4634.939578, -48865.402741, -29131.062007, 102033.891774, -83692.857560\right)$\\
        $B_4$ & $F_3\left(-108.617223, 51.647256, -80.701863, 100.737380, 599.399725, 4524.130261, 5732.389194, 63783.814535, 26674.381819, -93405.082867, 76825.313280\right)$\\
        \hline
        \multicolumn{2}{c}{Progenitor mass above MHeF}\\
        \hline
		Coefficient & Expression \\
		\hline
        $B_1$ & $F_1\left(22.025096, 0.181156, -0.091887, 2.907441, 61.720625, -186.899498, -64.117643, -1338.448779\right)$\\
        $B_2$ & $F_2\left(6.518631, -0.076714, -8.964896, -0.397295, 4.627599, -1034.098577, 3.034269, -3.262960, 1.071048\right)$\\
        $B_3$ & $F_3\left(-24.259990, 9.674470, 33.807359, -28.283298, 93.282366, 22218.746585, 254.576441, 1269.137159, -7318.853816, 13786.689347, -8458.441421\right)$\\
        $B_4$ & $F_3\left(7.879596, -5.387333, -5.251055, -58.010186, 557.107928, 44336.330592, -131.885596, -2088.395847, 4838.914927, -9979.402962, 6354.071627\right)$
         \botrule
	\end{tabular}}
\end{table}

\begin{table}
    \centering
	\caption{\centering Coefficients for the luminosity fit, as presented in Equation \ref{eq:luminosity}. The relevant $F_i$ expressions are defined in appendix \ref{ap:lum}.}\label{tab:C}
	{\tablefont\begin{tabular}{@{\extracolsep{\fill}}lc}
		\toprule
        \multicolumn{2}{c}{Progenitor mass below MHeF}\\
        \hline
		Coefficient & Expression \\
		\hline
        $C_1$ & $F_4\left(-2.832156, 1.893389, -8.330816, 0.362427, 2.673373, 1.079903, 8.669519\right)$\\
        $C_2$ & $F_4\left(4.347669, 11.368773, -51.446872, 3.973225, -33.695657, -9.207025, -424.934405\right)$\\
        $C_3$ & $F_4\left(2.038440, -42.792174, 96.225155, 1.838567, 13.540573, 5.336893, -7.124086\right)$\\
        $C_4$ & $F_5\left(0.493809, 2.582426, -1.595225, -0.299580, 0.5, -0.018246, -3.431202, 0.173840, -2.045155\right)$\\
        $C_5$ & $F_5\left(0.205285, -2.542241, 0.899499, -0.656101, 1, 0.008024, 0.276488, -0.486153, 0.546976\right)$\\
        $C_6$ & $F_6\left(-0.999444, 38.450766, -75.372175, 211.460780, 11.372078, -598.968156, -22.253736, 1481.736167\right)$\\
        \hline
        \multicolumn{2}{c}{Progenitor mass above MHeF}\\
        \hline
		Coefficient & Expression \\
		\hline
        $C_1$ & $F_4\left(1.290771, 0.011268, -1.551169, 1.432352, 46.903981, -1.659297, -59.940697\right)$\\
        $C_2$ & $F_4\left(0.845934, 1.062164, -44.727936, 3.382014, 110.443012, -21.708097, 56.817945\right)$\\
        $C_3$ & $F_4\left(1.805946, -18.760767, 43.447356, 4.459667, 354.682423, 0.673488, 271.354206\right)$\\
        $C_4$ & $F_5\left(-0.430926, -6.049646, -0.826707, -0.298135, 0.5, -0.063759, -2.422614, -0.142219, -54.654176\right)$\\
        $C_5$ & $F_5\left(1.095360, 0.033632, 27.629045, -28.972798, 1, 465.409144, 112269.817261, -66.539093, -15917.982883\right)$\\
        $C_6$ & $F_6\left(-0.819856, 36.965087, -46.603410, 115.884948, 5.492285, 218.203227, -10.880590, 129.648596\right)$
         \botrule
	\end{tabular}}
\end{table}

\end{landscape}

\begin{figure*}[ht!]
\centering
\includegraphics[width=1\linewidth]{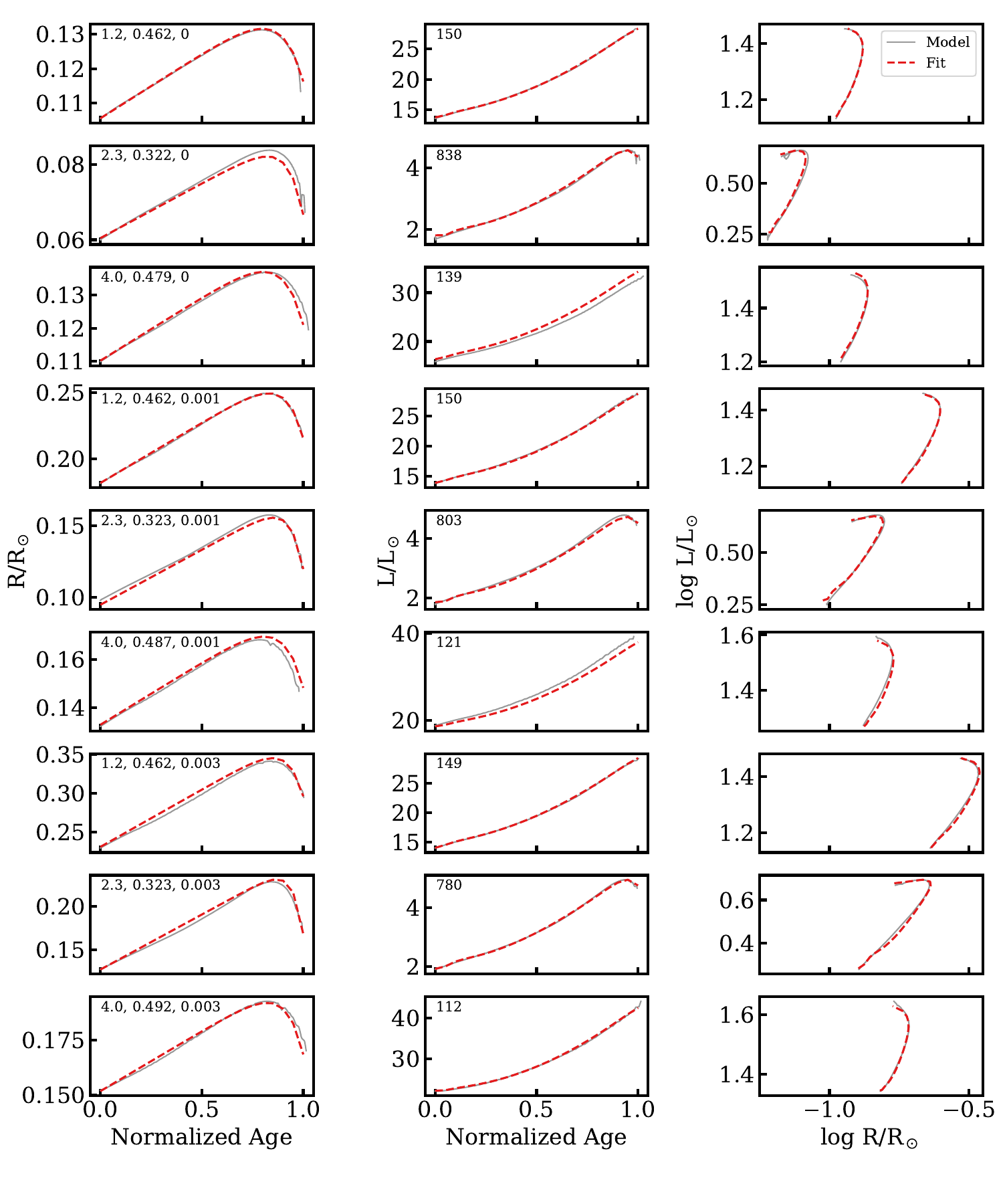}
\caption{Graphic comparison between the prescription developed in section \ref{sec:h_prescription} (red lines) and the models from \citet{Bauer2021} (solid grey lines). On the top-left corner of the left-hand side panels, information related to the properties of the HeMS model in each row is given in the format \textit{mass at ZAMS} (M$_\odot$), \textit{HeMS mass} (M$_\odot$), \textit{metallicity} (Z). A similar annotation in the middle panel shows the total time spent (Myr) in the HeMS stage.}
\label{fig:bauer_diff}
\end{figure*}

\end{appendix}
\end{document}